\documentclass[aps,pra,a4paper,reprint,showpacs,nofootinbib,superscriptaddress,floatfix,floats]{revtex4-1}
\usepackage{pifont}
\usepackage{amsmath}
\usepackage{amssymb}
\usepackage{amsfonts}
\usepackage{times,txfonts}
\usepackage{bm}
\usepackage{bbm}
\usepackage{dsfont}
\usepackage{bbold}
\usepackage{hyperref}
\hypersetup{
    colorlinks=true,
    citecolor=blue,
    linkcolor=blue,
    urlcolor=blue
}
\usepackage{color}
\usepackage{graphicx}
\usepackage{float}
\usepackage{epsfig}
\usepackage{verbatim}
\usepackage{stackengine} 

\newcommand{\ket}[1]{\left\vert#1\right\rangle}

\newcommand{\ketbra}[2]{\left|{#1}\rangle \langle{#2}\right|}
\newcommand{\inner}[2]{\left\langle#1\kern-\nulldelimiterspace\left|#2\kern-\nulldelimiterspace\right.\right\rangle}

\newcommand{\sech}{\mathrm{sech}}

\newcommand\matd[2]{\begin{bmatrix} #1 \\ #2 \end{bmatrix} }
\newcommand{\half}{\frac{1}{2}}

\newcommand{\etal}{{\it et al.}\ }
\newcommand{\ie}{i.e.}
\newcommand{\eg}{e.g.}
\def\stackalignment{l}

\begin{document}

\title{Quantum walk as a simulator of nonlinear dynamics: Nonlinear Dirac equation and solitons}

\author{Chang-Woo Lee}
\affiliation{School of Computational Sciences, Korea Institute for Advanced study, Dongdaemun-gu, Seoul 130-722, Korea}
\affiliation{Department of Physics, Texas A\&M University at Qatar, PO Box 23874, Doha, Qatar}
\author{Pawe\l{} Kurzy\'{n}ski}
\affiliation{Centre for Quantum Technologies, National University of Singapore, 3 Science Drive 2, 117543 Singapore, Singapore}
\affiliation{Faculty of Physics, Adam Mickiewicz University, Umultowska 85, 61-614 Pozna\'{n}, Poland}
\author{Hyunchul Nha}
\affiliation{School of Computational Sciences, Korea Institute for Advanced study, Dongdaemun-gu, Seoul 130-722, Korea}
\affiliation{Department of Physics, Texas A\&M University at Qatar, PO Box 23874, Doha, Qatar}

\date{\today}

\begin{abstract}
Quantum walk (QW) provides a versatile tool to study fundamental physics and also to make a variety of practical applications. We here start with the recent idea of {\it nonlinear} QW and show that introducing {\it nonlinearity} to QW can lead to a wealth of remarkable possibilities, e.g., simulating nonlinear quantum dynamics thus enhancing the applicability of QW above the existing level for a universal quantum simulator. As an illustration, we show that the dynamics of a nonlinear Dirac particle can be simulated on an optical nonlinear QW platform implemented with a measurement-based-feedforward scheme.
The nonlinear evolution induced by the feed-forward 
introduces a self-coupling mechanism to (otherwise linear) Dirac particles, which accordingly behave as a \emph{soliton}. We particularly consider two kinds of nonlinear Dirac equations, one with 
a scalar-type self-coupling (Gross-Neveu model) and the other with a vector-type one (Thirring model), respectively. Using their known stationary solutions, 
we confirm that our nonlinear QW framework is capable of exhibiting characteristic features of a soliton. 
Furthermore, we show that the nonlinear QW enables us to observe and control an enhancement and suppression of the ballistic diffusion.
\end{abstract}


\maketitle

\section{Introduction}
A random walk---a process of making a random choice among multiple paths---naturally arises in numerous situations. It not only provides an important conceptual basis in statistics \cite{RG04}, but also has a wide range of applications in various areas of physics \cite{Barber70}, computer science \cite{MR95}, and biology \cite{Berg93}. 
Its quantum mechanical analog, i.e., quantum walk (QW), was also proposed by Aharonov {\it et al.} \cite{Aharonov93}. As the QW draws on the principle of superposition, it can remarkably manifest a variety of novel features via quantum coherence beyond the classical random walk (CRW). 
For instance, the diffusion $\sigma$ of the quantum walker is much faster scaling as $\sigma_Q\sim t$ unlike the case of classical walker $\sigma_C\sim \sqrt{t}$, where $t$ represents the number of random steps. 
It can find a number of applications including fast search algorithms \cite{Farhi98,Shenvi03,Childs04}, graph isomorphism test \cite{Gamble10,Berry11,Wang15}, and boson sampling \cite{Franson13,Broome13,Spring13}.

QW was also proposed as a primitive for universal quantum computation \cite{Childs09,Childs13} and as a tool to simulate quantum dynamics
\cite{Strauch0605,Kurzynski08,Chandrashekar13,Molfetta13,Witthaut10,Preiss15}. 
Just as cellular automata can be adopted to simulate, and thus understand better, various dynamical systems \cite{ZG88}, QW can provide a crucial tool as a quantum cellular automaton to address quantum dynamical systems \cite{Kempe03,Venegas12}. 
QW has been experimentally implemented in a number of physical systems including neutral atoms in optical lattice \cite{Karski09,Preiss15}, trapped ions \cite{Schmitz09,Zahringer10}, and photons \cite{Schreiber10,Schreiber12,Jeong13}, extending to multiparticle \cite{Preiss15} and multidimensional cases \cite{Schreiber12,Jeong13}. 

In this article, we explore a possibility of exploiting QW to a larger extent, i.e., to use QW as a simulator of {\it nonlinear} quantum dynamics, thus demonstrating a much enhanced applicability of QW beyond the existing level. 
It is well understood that a nonlinearity can be implemented onto a dynamical system, e.g., by a measurement-based feedforward scheme, 
which affects the evolution of a quantum system conditioned on the measurement outcome at previous steps. 
Recently, Shikano {\it et al.} studied a nonlinear QW via feed-forward scheme \cite{Shikano14}, where the investigation was, 
however, limited to the emergence of anomalous slow diffusion of QW scaling as $\sigma_C\sim t^{0.4}$. 
In contrast, we show here that there exist a wider range of applications of nonlinear QW and illustrate that it can be used to simulate the dynamics of nonlinear Dirac particles. 
While many studies in linear QW so far focused on the diffusive dynamics leading to delocalized wave function, 
we here demonstrate a solitonic behavior of nonlinear Dirac particle and even the coherent collisions of solitons manifesting the quantum nature of coherence. 
Furthermore, we also demonstrate that the diffusion speed of quantum walker can be widely tunable in the framework of nonlinear QW. 
We anticipate that a richer set of dynamical features can also emerge in a multidimensional nonlinear QW.

\section{Quantum Walk and Dirac equation}
Several works so far considered QW to emulate a relativistic fermion like a free Dirac particle \cite{Strauch0605,Bracken07,Kurzynski08,Chandrashekar10,Chandrashekar13,Molfetta13},  which we brifely review here.
We only consider 1+1-dimensional (one temporal and one spatial) Dirac equation (DE), while 1+3-dimensional DE can also be effectively simulated on the QW platform \cite{Chandrashekar13}.
A 1+1 dimensional DE can be written as
$ (i \gamma^\mu \partial_\mu - m c)\psi = 0 ,$
where 
$ \partial_\mu = (\partial_0, \partial_1)  = (\partial_{ct}, \partial_x) $ and the exact form of Dirac matrices $\gamma^\mu$ is shown later.
We take $\hbar = 1$ and the DE can be put in the Hamiltonian form as
\begin{equation}
i \partial_t \psi = (- i c \alpha \partial_x  + \beta m c^2)\, \psi ,
\end{equation}
where the Hermitian matrices $ \alpha$ and $\beta $ must satisfy
\begin{equation}
\alpha \beta + \beta \alpha = 0,\quad \alpha^2 = \beta^2 = \mathds{1}.
\label{eq:rel_alpha_beta}
\end{equation}

The dynamics of DE can be efficiently realized on a QW platform, which
consists of two operations, coin-flipping and shift.
A coin operator coherently mixes a walker's internal states, 
\eg, spin up ($\uparrow$) and down ($\downarrow$) states for a two-dimensional case.
A general form of a two-dimensional coin operator is the SU(2) operator 
and it suffices here to employ its simple form as
\begin{equation}
\hat C = \matd{\cos \Theta & - \sin \Theta}{\sin \Theta & \phantom{-} \cos \Theta} .
\label{eq:coin}
\end{equation}
A shift operator moves a walker by one step left or right in position space according to its internal state 
\begin{equation}
\hat S = \sum_{x} \big[ \ketbra{x+1}{x} \otimes \ketbra{\uparrow}{\uparrow} + \ketbra{x-1}{x} \otimes \ketbra{\downarrow}{\downarrow} \big] .
\end{equation}
If we define a displacement operator using momentum $ \hat p $ as 
$ e^{\pm i\hat p \Delta x} \psi (x) = \psi (x \pm \Delta x)$,
the shift operator has the matrix representation in coin space as $\hat S = \matd {e^{+i \hat p} &  0}{0 & e^{-i\hat{p}}}$. 
Thus, the dynamics of a walker is determined by the combined evolution
\begin{equation}
\hat W_\mathrm{L} 
= \hat S \hat C
= \matd
  {\cos\Theta \,e^{+i \hat p} &  -\sin \Theta \,e^{+i\hat{p}}}
  {\sin \Theta \,e^{-i\hat{p}} & \phantom{-} \cos \Theta \,e^{-i\hat{p}}},
\label{eq:w_linear}
\end{equation}
where the subscript $L$ refers to linear QW.
If we represent the state of a walker at time $t$ as
\begin{eqnarray}
\ket{\psi(t)} 
= \sum_{x} \ket{x} \otimes \left[u(t,x) \ket{\uparrow} + v(t,x) \ket{\downarrow} \right]
\equiv \sum_{x} \ket{x} \otimes \matd{u(t,x)}{v(t,x)},\nonumber \\
\end{eqnarray}
with spin-dependent wavefunctions $u(t,x)$ and $v(t,x)$, respectively, 
the state after one step through Eq. \eqref{eq:w_linear}, i.e.,
$ \ket{\psi(t+1)} = \hat W_\mathrm{L} \ket{\psi(t)}$, is given by the relation
\begin{equation}
\matd{u(t+1,x+1)}{v(t+1,x-1)} = \hat C \matd{u(t,x)}{v(t,x)}.
\label{eq:evol_uv_linear}
\end{equation}
Following this evolution sequentially,
we obtain a state at time $ t$ from an initial state $\ket{\psi(0)}$ as 
$\ket{\psi(t)} = \hat W_\mathrm{L}^t \ket{\psi(0)}$.
Mapping of discrete QW onto DE have been considered, e.g., by Strauch \cite{Strauch0605} and 
Chandrashekar \cite{Chandrashekar13}, and we adopt Chandrashekar's formalism as follows.

To obtain the desired effective Hamiltonian of QW, 
Chandrashekar \cite{Chandrashekar13} takes the logarithm of 
$ \hat W \equiv e^{-i \hat H_\mathrm{L} \Delta t}$ ($\Delta t = 1$); \ie,
\begin{equation}
-i \hat H_\mathrm{L} = \ln \hat W_\mathrm{L} = \hat V (\ln \hat D) \hat V^{-1}.
\end{equation}
Here 
$ \hat D = \mathrm{diag} (e^{+i \hat \omega},~e^{-i \hat \omega} ) $ 
with 
\begin{equation}
e^{\pm i \hat \omega} = \cos\Theta \cos \hat p \pm i \sqrt{1- \cos^2 \Theta \cos^2 \hat p} 
\label{eq:omega}
\end{equation}
and the exact form of the matrix $\hat V$ is detailed in Ref. \cite{Chandrashekar13}.
After some algebra one can obtain the following effective Hamiltonian:
\begin{equation}
\hat H_\mathrm{L} = \frac{\hat \omega}{\sin \hat \omega} \matd
{-\cos\Theta \sin \hat p & i \sin \Theta e^{+i\hat p}}
{i \sin \Theta e^{-i\hat p} & -\cos\Theta \sin \hat p } \cdot \sigma_3 ,
\end{equation}
where $\sigma_i~(i=1,2,3)$ are Pauli matrices. 
The operator $ \sin \hat p $ may be understood by way of exponential operators $e^{\pm i\hat p}$ as
\begin{eqnarray}
\sin \hat p ~ \psi (t,x) 
&=& \frac{1}{2i} \left( e^{+i\hat p} - e^{-i\hat p}\right) \psi (t,x) \nonumber\\
&=& \frac{1}{2i} \left[\psi (t,x+1)  - \psi (t,x-1) \right] \nonumber\\
&\approx& -i \frac{\partial}{\partial x} \psi (t,x) .
\end{eqnarray}
The operator $e^{\pm i\hat p}$ can also be replaced by its differential form as
\begin{eqnarray}
e^{\pm i\hat p} \psi (t,x) 
&=& \psi (t,x \pm 1) 
= \psi (t,x \pm 1) - \psi (t,x) + \psi (t,x) \nonumber\\
&\approx& \left( \pm \frac{\partial}{\partial x} + 1 \right) \psi (t,x) .
\end{eqnarray}
We thus have the operators in approximate forms 
\begin{equation}
\sin \hat p \approx -i \frac{\partial}{\partial x}, \quad
e^{\pm i\hat p} \approx \left( 1 \pm \frac{\partial}{\partial x} \right),
\end{equation}
and $ \cos \hat p \approx 1 $, which is valid to the first-order in momentum operator $\hat p$.
Taking 
$\hat \omega / \sin \hat \omega \approx 1/ \cos \hat \omega \approx 1/ \cos \Theta$ 
in view of l'Hosptial's rule,  
the Hamiltonian of QW becomes that of a Dirac particle, \ie,
\begin{eqnarray}
\hat H_\mathrm{L}
&=& - i \sec \Theta \matd{-\cos \Theta & \sin \Theta}{ \sin \Theta & \cos \Theta} \frac{\partial}{\partial x} + \matd{0 & -i}{i & 0} \tan\Theta \nonumber\\
&=& - i c \alpha \partial_x  + \beta m c^2,
\label{eq:h_eff}
\end{eqnarray}
with the identifications
\begin{equation}
c = \sec \Theta,  \quad m c^2 = \tan \Theta,
\label{eq:m_c}
\end{equation}
and 
\begin{equation}
\alpha = -\cos\Theta \,\sigma_3 + \sin\Theta \,\sigma_1, \quad
\beta = \sigma_2 .
\label{eq:alpha_beta_rot}
\end{equation}
Note that Chandrashekar's approach takes $\hat \omega / \sin \hat \omega \approx 1$,
but we find that the parametrization $\hat \omega / \sin \hat \omega \approx 1/ \cos \Theta$ gives better results.
It is readily seen that the operators in Eq. \eqref{eq:alpha_beta_rot} satisfy the conditions in Eq. \eqref{eq:rel_alpha_beta} 
required for the dynamics of Dirac particle.

\section{Nonlinear Dirac Equation on Quantum Walk Platform}
\subsection{Nonlinear Dirac equation}
We here move forward to address a possibility of simulating 
a \emph{nonlinear} Dirac particle on the QW platform.
Nonlinear Dirac equation (NDE) is regarded as a model describing extended particles without invoking quantization or intermediate bosons (gluon) mediating strong force as in 
the quark-gluon model of quantum chromodynamics \cite{Kaus76,Ranada84}. In particular, 
NDE can well describe hadrons as its particlelike solutions and
strong interactions by means of nonlinear self-coupling.
Just like other nonlinear equations that can have a solitonic solution, 
NDE also allows solitary-wave (or particlelike) solutions---stable localized solutions with finite energy and charge---and 
those can be boosted to have an arbitrary velocity and remain particlelike even after collision with each other.
There are various NDEs classified according to interaction and topological types. We here consider nontopological scalar and vector-type interactions, i.e.,
massive Gross-Neveu model \cite{Gross74} and Thirring model \cite{Thirring58}.
The exact analytical solutions for these models are known 
\cite{Lee75,Alvarez81,Stubbe86,Blanchard87,Tran14}, which can be used to test our discrete-time QW framework as an NDE simulator.

Let us begin with Lagrangian density with a generic nonlinear self-coupling term as
\begin{equation}
\mathcal{L} = \frac{i c}{2} \left[ 
\bar{\psi} \gamma^\mu \partial_\mu \psi - 
(\partial_\mu \bar{\psi}) \gamma^\mu \psi
\right] 
- m c^2 \bar{\psi} \psi 
+ G (\bar{\psi} \Gamma \psi) ,
\label{eq:lagrangedensity}
\end{equation}
where $\bar{\psi} = \psi^\dag \gamma^0$ and $G$ is a real-valued function with $G(0)=G'(0)=0$.
The matrix $ \Gamma $ determines the type of interaction, 
\ie, $ \Gamma = \mathds{1} ~(\gamma_\mu) $ for scalar (vector) type.
For brevity, let us adopt the representation of the $\gamma$ matrices as
\begin{equation}
\gamma^0 = \sigma_3,~ \gamma^1 = i \sigma_1~(\text{namely,}~~ \beta = \sigma_3, ~ \alpha = - \sigma_2).
\label{eq:gammamatrix2}
\end{equation}
We later transform by rotation the above matrices 
so that they match the specific form in Eq. \eqref{eq:alpha_beta_rot}.
We obtain the equations of motion from the Lagrange density in Eq. \eqref{eq:lagrangedensity} as
\begin{equation}
i c \gamma^\mu \partial_\mu \psi - m c^2 \psi + \frac{\partial G}{\partial \bar{\psi}} = 0, 
\label{eq:nlde1}
\end{equation}
which, for the specific case of $ G(x) = \half g x^2$, \ie, Gross-Neveu and Thirring model, is reduced to a Hamiltonian form
\begin{equation}
i \partial_t \psi = \hat H_\mathrm{NL} \psi \equiv [\hat H_\mathrm{L} + \hat h_\mathrm{NL} (\psi)] \psi.
\label{eq:h_nl}
\end{equation}
Here, 
the state-dependent nonlinear part
$\hat h_\mathrm{NL} (\psi) = - g(\bar{\psi} \Gamma \psi)\ \beta \Gamma$
becomes
\begin{equation}
\hat h_\mathrm{NL} (\psi) = - g(\bar{\psi} \psi) \,\beta
\label{eq:h_nl_scalar}
\end{equation}
for scalar-type interaction (Gross-Neveu model) and
\begin{equation}
\hat h_\mathrm{NL} (\psi) = - g(\psi^\dag \psi) ,
\label{eq:h_nl_vector}
\end{equation}
for vector type (Thirring model), respectively.

Now seeking stationary solutions of the form
\begin{equation}
\psi_\text{st} (t,x) = e^{-i \omega t} \varphi_\text{st} (x) = e^{-i \omega t} \matd{u_\text{st}(x)}{v_\text{st}(x)}
\end{equation}
particularly a real solution for position dependent part, one obtains \cite{Lee75,Alvarez81,Stubbe86,Blanchard87}
\begin{subequations}
	\begin{eqnarray}
	u_\text{st}(x) &=& \sqrt{\frac{2(mc^2-\omega)}{g}} \frac{\sech (b x)}{1 \mp a^2 \tanh^2(b x)}, \label{eq:stationary_sol_u}\\ 
	v_\text{st}(x) &=& a \tanh(b x) \, u_\text{st}(x) ,
	\label{eq:stationary_sol_v}
	\end{eqnarray}
	\label{eq:stationary_sol}
\end{subequations}
where 
$ a = \sqrt{(mc^2-\omega)/(mc^2+\omega)}, ~ b = \sqrt{m^2 c^4-\omega^2} $,
and the ``--'' (``+'') sign in Eq. \eqref{eq:stationary_sol_u} 
corresponds to the scalar (vector) type interaction.
For completeness,
an analytical solution for pseudoscalar-type interaction
$\Gamma =  \gamma^5 = i \gamma_0 \gamma_1$ \cite{Lee75,Stubbe86} appears as
\begin{equation}
u_\text{st}(x) = \sqrt{\frac{2(mc^2-\omega)}{g}} \frac{1 + a^2 \tanh^2(b x)}{2 a \sinh (b x)} ,
\end{equation}
with $ v_\text{st}(x) $ being the same form as Eq. \eqref{eq:stationary_sol_v}.
However, it is not normalizable and presumably not feasible in our setup, so we do not consider it here.

We may define the \textit{charge} as 
\begin{equation}
Q = \int \psi_\text{st}^\dag \psi_\text{st}^{} \,dx ,
\label{eq:charge}
\end{equation}
which is reduced to
\begin{equation}
Q = \frac{2 b}{g \omega}
\end{equation}
and
\begin{equation}
Q = \frac{2}{g} \tan^{-1} \frac{b}{\omega} ,
\end{equation}
for scalar and vector interactions, respectively \cite{Blanchard87}.

\begin{figure*}[t]
	\includegraphics[width=2\columnwidth]{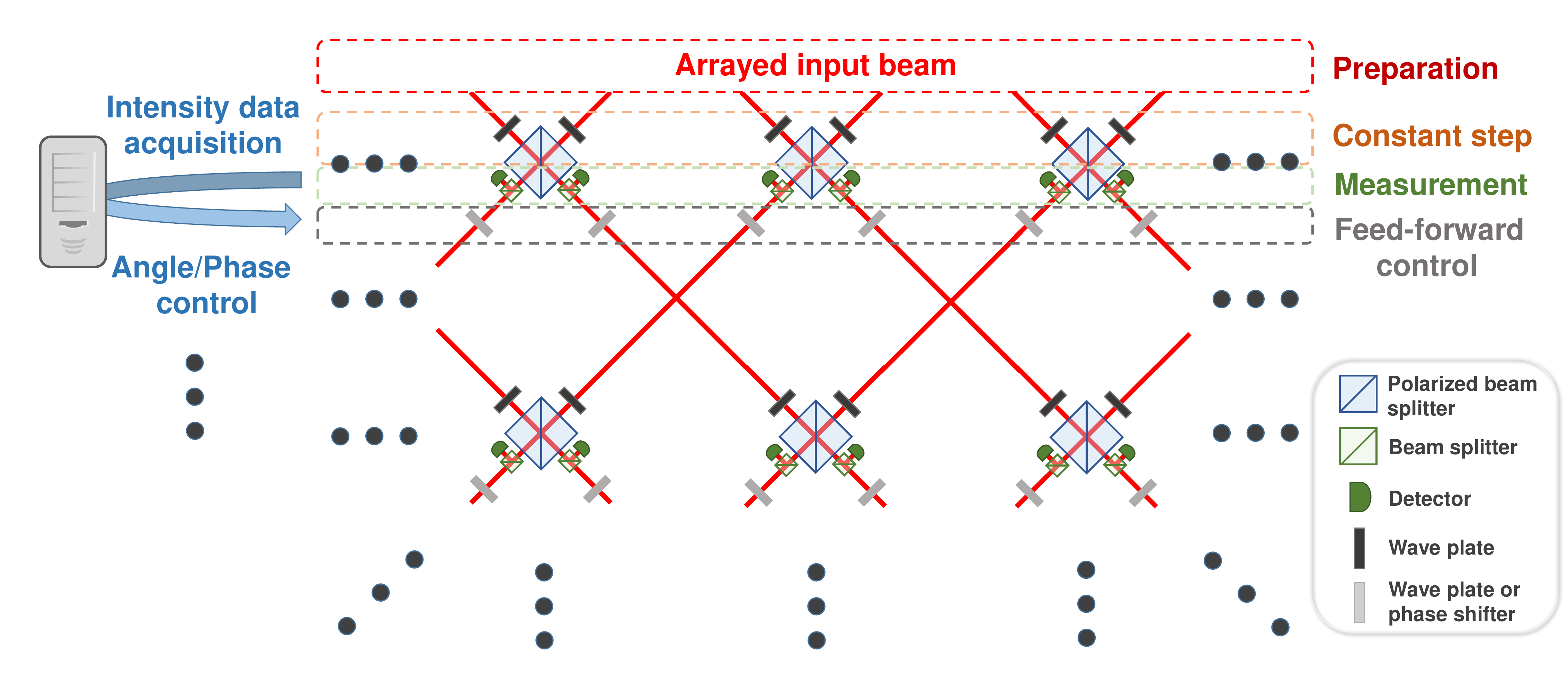}
	\caption{(Color online)
		Optical setup for the nonlinear quantum walk (NQW) using feed-forward scheme.
		The schematic shown here is part of the entire setting and 
		the blobs imply that the same components repeat in the corresponding directions.
		The polarization of each beam is rotated by a wave plate implementing the coin operator in Eq. \eqref{eq:w_linear}  and
		the beam is split into two by a polarized beam splitter (shift operator), which implements the linear QW evolution.
		Each WP in the figure is actually a combination of quarter and half WPs
		but we refer to it collectively by WP for brevity. On the other hand, 
		the intensity of each split beam is measured and used to control the succeeding WP (scalar-type NQW) or phase shifter (vector-type NQW), respectively, which generates a nonlinear evolution.
	}
	\label{fig:setup}
\end{figure*}

\subsection{Quantum walk simulating nonlinear Dirac equation}

Interestingly, although the model describes a quantum particle, its main features  can be simulated with classical light.
Using polarized light beams (walker), wave plates (coin operator), and polarized beam splitters (shift operator), 
one can realize a QW in an optical system.
In addition to this linear QW scheme, 
Shikano \etal \cite{Shikano14} recently studied the case of including an intensity measurement between polarized beam splitters (PBSs) and wave plates (WPs). Conditioned on the measured intensity, the angle parameters of the next-step WPs may be adjusted, \ie, the subsequent coin operators depend on the previous measurement outcomes.
As a result, 
they obtained a specific nonlinear equation for QW, i.e. \textit{porous medium equation}, 
$\frac{\partial}{\partial t} p(t,x) = \frac{\partial^2}{\partial x^2} p^{m}(t,x).$ 
They investigated an angle control scheme to result in $ m = 3/2 $, 
showing an anomalous diffusion $\sigma_m (t) \propto t^{2/5}$.

In our scheme, slightly different from the one in Ref. \cite{Shikano14}, we instead control  
\textit{ additional} WPs (scalar-type) or phase shifters (vector-type) conditioned on the measured intensity, 
which are placed between the previous-step PBSs  
and the next-step WPs
in Fig. \ref{fig:setup}. We measure the coin-state-dependent intensities $|u(x)|^2$ and $|v(x)|^2$ at each localtion $x$, as we need information on $\bar \psi \psi=|u(x)|^2-|v(x)|^2$ $\left(\psi^\dag \psi=|u(x)|^2+|v(x)|^2\right)$ for a scalar (vector)-type interaction.
The coin angle in ordinary WPs must not be changed for the particle mass $m$ and matrix $\alpha$ as shown in Eqs. \eqref{eq:m_c} and \eqref{eq:alpha_beta_rot}.
As a result, we obtain a \textit{nonlinear} evolution operator by including
$e^{-i \hat h_\mathrm{NL}}$ [Eqs. \eqref{eq:h_nl_scalar} and \eqref{eq:h_nl_vector}] 
after the linear evolution 
$\hat W_\mathrm{L} = e^{-i \hat H_\mathrm{L}}$ [Eqs. \eqref{eq:w_linear} and \eqref{eq:h_eff}], i.e.,
\begin{equation}
\hat W_\textrm{NL} = e^{-i \hat h_\mathrm{NL}} \hat W_\textrm{L} 
= e^{-i \hat h_\mathrm{NL}}  e^{-i \hat H_\mathrm{L}}
\approx e^{-i \hat H_\textrm{NL}}
\label{eq:w_nl}
\end{equation}
with 
$\hat H_\textrm{NL} = \hat H_\textrm{L}+ \hat h_\textrm{NL}$ as defined in Eq. \eqref{eq:h_nl} \cite{note}. 
In order to realize the above evolution on the optical setup in Fig. \ref{fig:setup},
the value of $\psi(t)$ in $\hat h_\textrm{NL} (\psi)$ is sampled at each position $x$ 
and used to compute $\psi (t+1)$.
Accordingly, Eq. \eqref{eq:evol_uv_linear} is modified to 
\begin{equation}
\matd{u(t+1,x+1)}{v(t+1,x-1)} = e^{-i \hat h_\mathrm{NL} [\psi(t,x)]} \hat C \matd{u(t,x)}{v(t,x)},
\label{eq:evol_uv_nonlinear}
\end{equation}
where   
$e^{-i \hat h_\mathrm{NL} [\psi(t,x)]}$ is just a phase 
for the vector-type case
while it is an  
operator for the scalar-type case that has the same form as the coin operator in Eq. \eqref{eq:coin} with $ \Theta = - g (\bar \psi \psi )$.

\begin{figure}[t]
	\topinset
	{\bfseries(a)}
	{\includegraphics[width=0.9\columnwidth]{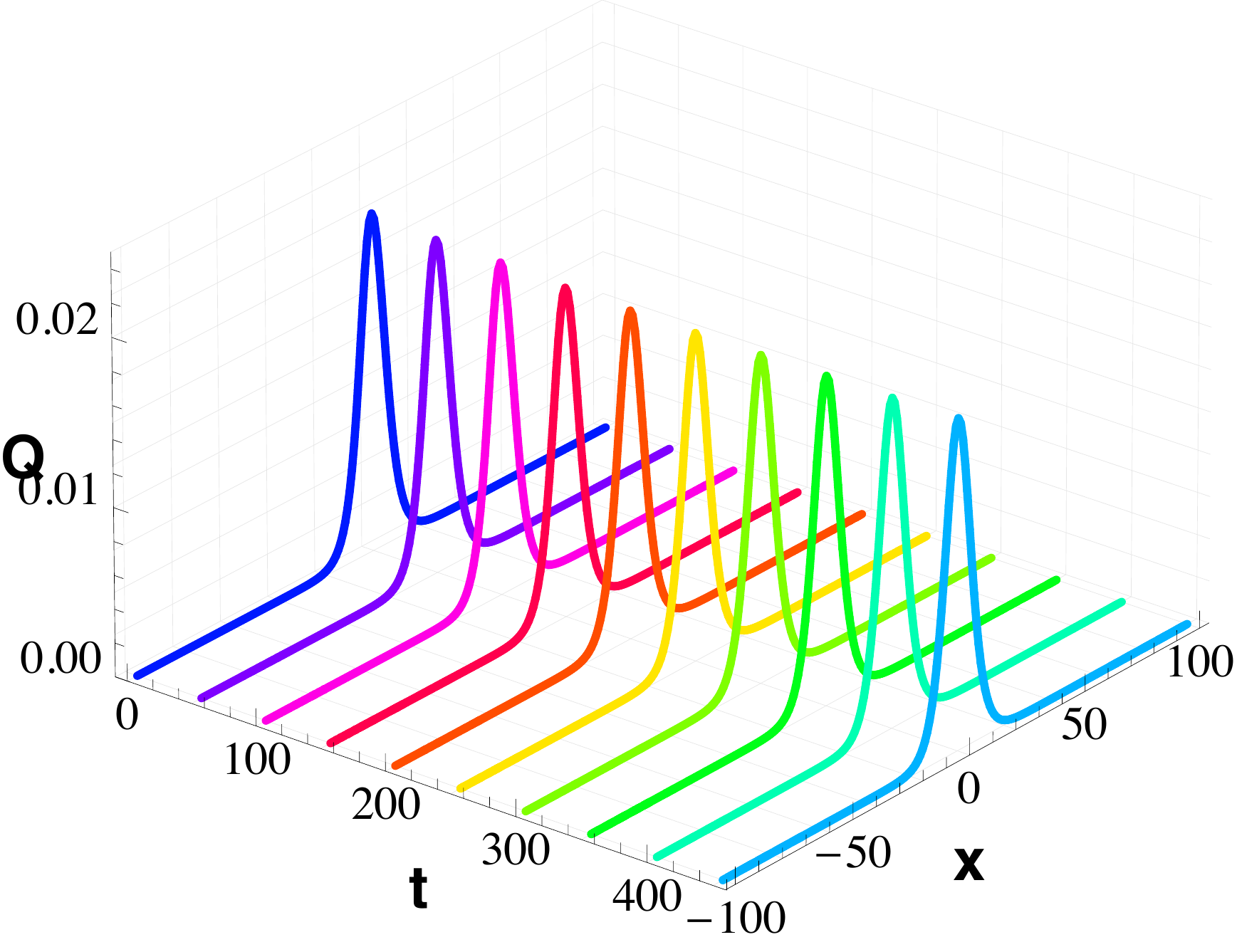}}
	{1mm}{1mm}
	\vskip1mm
	\topinset
	{\bfseries(b)}
	{\includegraphics[width=0.9\columnwidth]{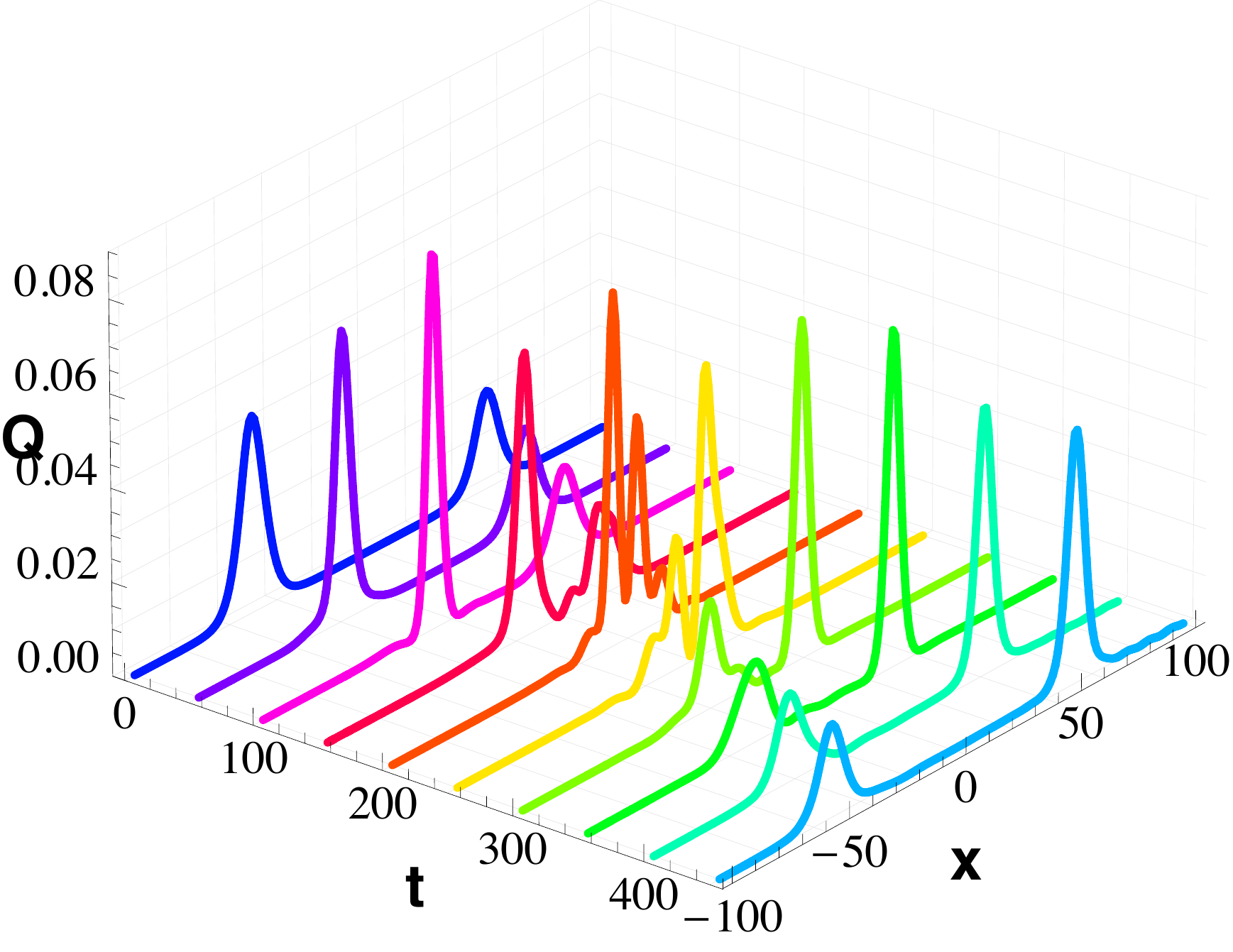}}
	{1mm}{1mm}
	\caption{(Color online)
		(a) Charge $Q$ (intensity distribution) of a stationary solution of NDE on NQW platform for scalar-type self-interaction.
		(b) Charge $Q$ of the superposition of oppositely moving solutions 
		on NQW platform for scalar-type self-interaction.
	}
\label{fig:soliton}
\end{figure}

\section{Solitonic Behavior and Controlling Ballistic Diffusion Speed}
We are now in a position to test if 
our NQW system can actually manifest a solitonic behavior, \ie, permanent particlelike stationary state.
To address this issue, 
we examine if the stationary solutions known for NDEs 
 are stationary on our discrete-time NQW platform as well.  
We thus prepare an initial state as 
$\ket{\psi_\mathrm{init}} = \sum_j \ket{j} \otimes R (\Theta) \matd{u_\text{st} (x=j)}{v_\text{st} (x=j)},$
where the values of 
$ u_\text{st}(x) $ and $ v_\text{st}(x) $ from Eq. \eqref{eq:stationary_sol} are sampled at discrete $ x=j $.
Note that the rotation matrix $R (\Theta) = R_y \left( \pi - \Theta \right) R_x \left( -\pi/2 \right)$,
with $R_k (\Theta)$ the rotation about $k$ axis, 
transforms the matrices 
$\alpha,~ \beta$ in Eq. \eqref{eq:gammamatrix2} to those in Eq. \eqref{eq:alpha_beta_rot}.
We compute the evolution of $\ket{\psi_\mathrm{init}}$ according to Eq. \eqref{eq:evol_uv_nonlinear} and 
in Fig. \ref{fig:soliton}(a),
we display its charge $Q$ (intensity in an optical setup) with respect to time and position for the case of scalar-type evolution. The evolution of vector-type case is similar, thus not shown here.
The self-coupling constant $g$ is set to unity and 
the energy $\omega$ to $0.99m$, so that the inverse width $b$ is small and 
consequently the profile has a modest width. 

Figure \ref{fig:soliton} shows that the charge never escapes from a finite region 
and the profile keeps almost the initial shape in a stable manner.
This dynamic stability or robustness is a well-known solitonic behavior, which was discussed for NDE \cite{Alvarez83}.
We thus confirm that 
even though the initial states above are not analytical stationary solutions in our NQW model,
they definitely show a solitary and stable solitonic behavior with no transport. In a QW system,
localization can also be induced, e.g., by a spatial \textit{disorder} as shown in Refs. \cite{Keating07,Yin08}.
However, 
the localization in our case has a different physical origin, i.e., nonlinear self-interaction here indicated as an alternative localization mechanism.

One of the salient features of a soliton---and sometimes considered one of its requisites to be so called \cite{Drazin89}---is that 
its particlelike amplitude profile be maintained even after colliding with another soliton.
A Dirac soliton also has this unique property \cite{Alvarez81,Xu13} and
our NQW platform can successfully demonstrate this feature as well. 
To investigate this solitonic collision,
we must be equipped with the wave function of a \textit{moving} Dirac soliton.
We make use of the fact that a moving solution of Eq. \eqref{eq:nlde1} can be obtained by boosting a stationary solution in Eq. \eqref{eq:stationary_sol} with Lorentz transformations $x' = \gamma (x- v t)$ and $\quad t' = \gamma (t- v x / c^2),$
where 
$ \gamma = 1/\sqrt{1-(v/c)^2} $ is the Lorentz factor.
A relevant transformation matrix $\Lambda$ is given \cite{Wachter10} by
\begin{equation}
\Lambda (v) = \sqrt{\frac{\gamma+1}{2}} \,\mathds{1} - \frac{v}{|v|}\sqrt{\frac{\gamma-1}{2}} \,\alpha ,
\end{equation}
which gives the transformation of the wave function $\psi' (t',x') = \Lambda (v) \, \psi (t,x)$. 
Therefore, a traveling solution  
$ \psi_{v} (t,x)$ with a velocity $v$ 
is obtained from a stationary one $\psi_\mathrm{st} (t',x')$ by
\begin{equation}
\psi_{v} (t,x) \equiv \Lambda^{-1} (v) \, \psi_\mathrm{st} (t',x') .
\label{eq:moving_sol}
\end{equation}
We have checked that the velocity $v$ in our NQW system is allowed up to $0.8c$.
Beyond those values,
the states in Eq. \eqref{eq:moving_sol} become unstable and cannot maintain a sole peak profile.
We study collisional phenomena of two (or more) Dirac solitons 
by superposing oppositely moving solitons, i.e., an initial state $
\psi_\mathrm{init} (x) = c_+ \psi_{v} (x + x_0,t=0) + c_- \psi_{-v} (x - x_0,t=0).$
In Fig. \ref{fig:soliton}(b), 
we plot the time evolution of $\psi_\mathrm{init} (x)$ ($c_+ = \sqrt{2}$, $c_- = 1$, and $v=0.3c$.) 
for a scalar-type collision.
The two peaks representing independent solitons emerge again after transient behavior
and stabilize into almost the same shapes as the original ones, implying the maintenance of quantum coherence in our NQW framework.

\begin{figure}[!htb]
	\topinset
	{\bfseries(a)}
	{\includegraphics[width=0.45\columnwidth]{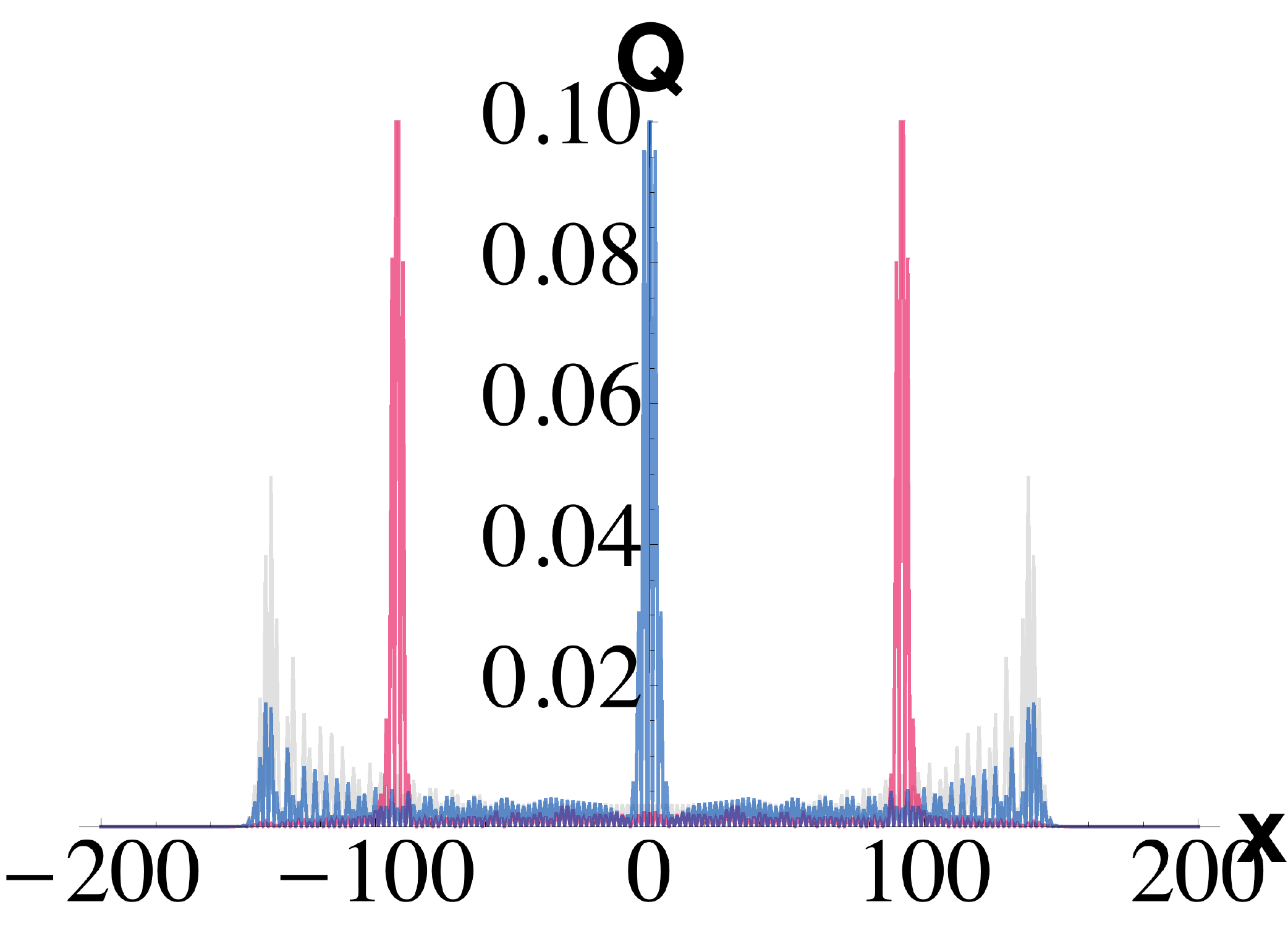}}
	{1mm}{-2mm}
	\hskip5mm
	\topinset
	{\bfseries(b)}
	{\includegraphics[width=0.45\columnwidth]{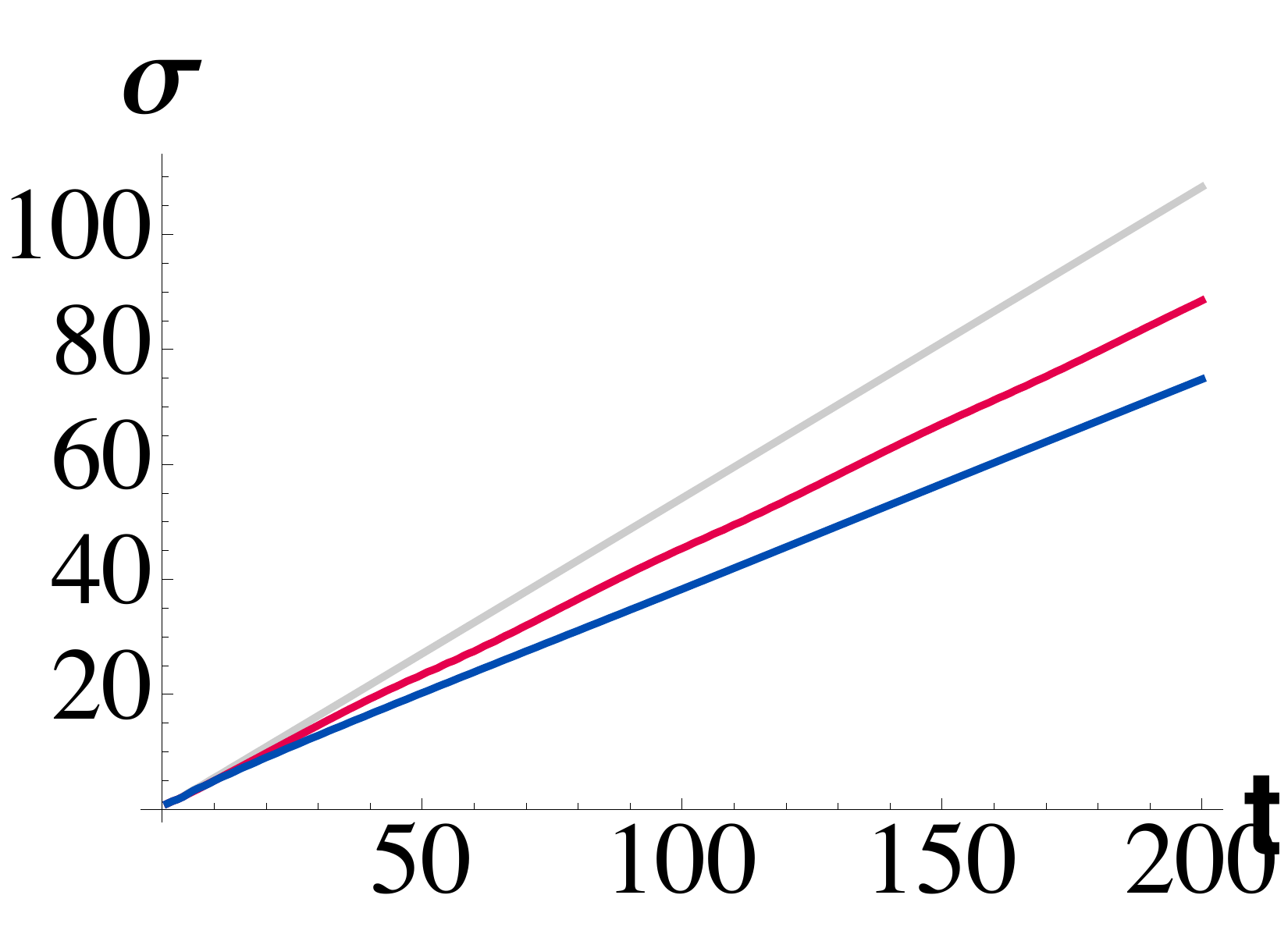}}
	{1mm}{35mm}
	\vskip1mm
	\topinset
	{\bfseries(c)}
	{\includegraphics[width=0.45\columnwidth]{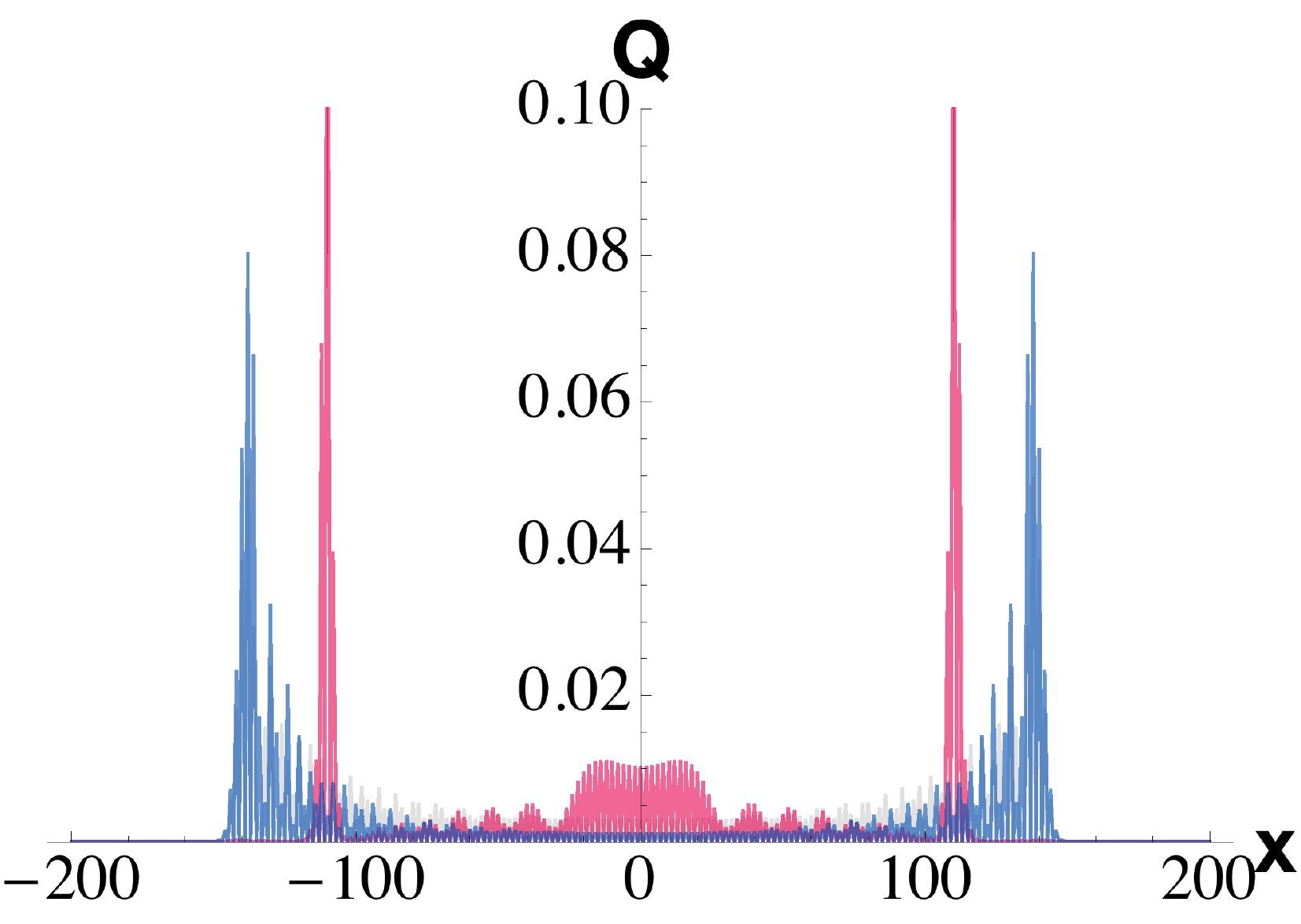}}
	{1mm}{-2mm}
	\hskip5mm
	\topinset
	{\bfseries(d)}
	{\includegraphics[width=0.45\columnwidth]{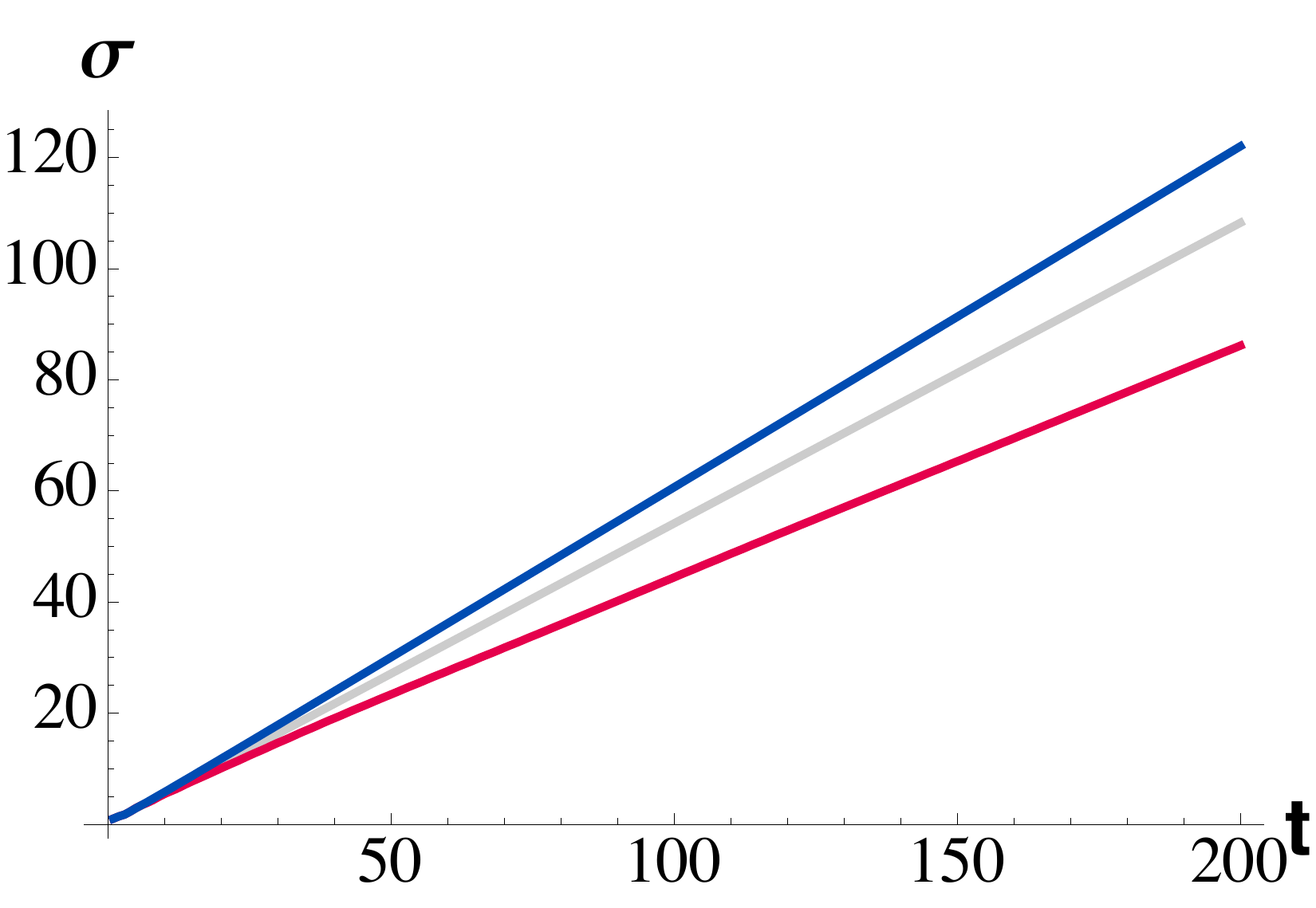}}
	{1mm}{35mm}
	\vskip1mm
	\hskip30mm
	\topinset	
	{\bfseries(e)}
	{\includegraphics[width=0.9\columnwidth]{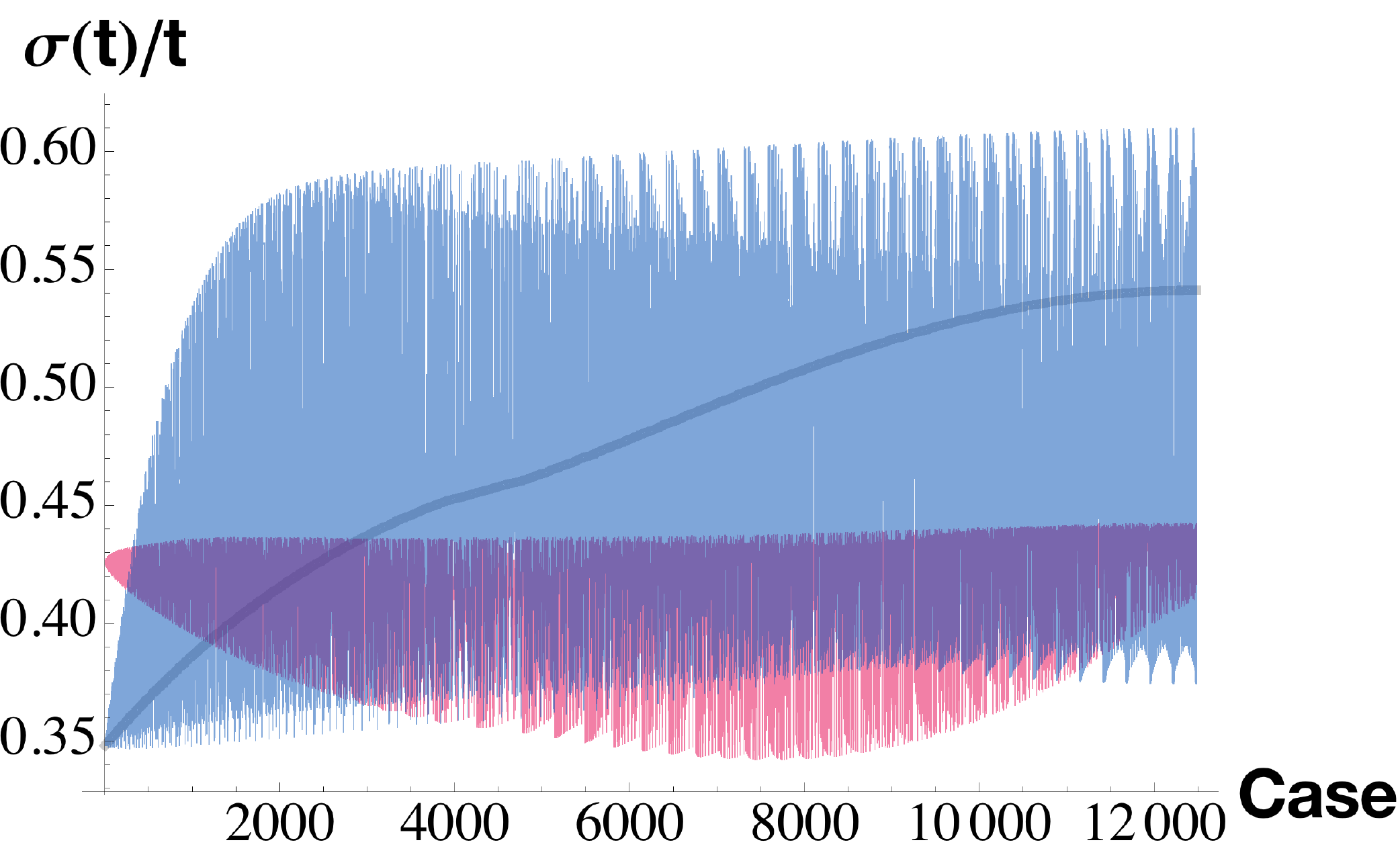}}
	{1mm}{-5mm}
	\caption{(Color online)
		(a)--(d) Snapshot of the charge distribution at $t=200$ of a walker starting from the origin with coin configuration
		$\ket{\varphi_+} = \ket{\uparrow} + i \ket{\downarrow}$ (a)  
		[$\ket{\varphi_-} = \ket{\uparrow} - i \ket{\downarrow}$ (c)] and 
		the corresponding growth of the position width $\sigma(t)$ with time (b) [(d)] until $t=200$.
		(e) Ballistic diffusion speed $\sigma(t)/t$ of linear QW, scalar-type NQW, and vector-type NQW 
		for 12,482 initial states of the form 
		$\ket{x=0} \otimes [\cos (\theta/2) \ket{\uparrow} + \sin (\theta/2) e^{i\phi} \ket{\downarrow}]$. 
		Note that the data points are arranged in ascending order of the speed value of linear QW for comparison.
		The gray curve denotes the case of linear QW, the blue curve the case of scalar type, 
		and the red curve the case of vector-type NQW.
	}
	\label{fig:diff_speed}
\end{figure}

We have so far demonstrated that 
our NQW scheme can provide a good simulation platform for a nonlinear Dirac particle.
NQW can give a deep insight into nonlinear quantum dynamics and also be further used for other practical applications. 
We here manifest one such interesting aspect of NQW, i.e., observation of a wide range of ballistic diffusion speed. 
Depending on interaction type and initial condition,
the ballistic speed can be significantly varied. That is, 
the particlelike behavior can speed up or slow down the spread of wave amplitude in a wide scope. 
Figure \ref{fig:diff_speed} shows the time evolution of an initial state 
$\ket{\psi_\mathrm{init}} = \ket{x=0} \otimes \ket{\varphi_\pm}$,
with the result for the coin state 
$\ket{\varphi_+} \equiv \ket{\uparrow} + i \ket{\downarrow}$ ($\ket{\varphi_-} \equiv \ket{\uparrow} - i \ket{\downarrow}$) 
shown in Figs. \ref{fig:diff_speed}(a) and \ref{fig:diff_speed}(b) [\ref{fig:diff_speed}(c) and \ref{fig:diff_speed}(d)] sampled at $t=200$.
As can be seen from Fig. \ref{fig:diff_speed}(a),
the case of initial coin state $\ket{\varphi_+}$ is contrasted with linear QW case 
(the gray plot in the figure). 
Although some portions travel away,
the charge is mostly concentrated in a certain finite region---in particular, around the origin in the case of scalar type.
This is owing to the localization property induced by self-interaction and 
accordingly the diffusion $\sigma(t)$ has a lower value for both interaction types than that of linear QW.
Notice, however, that its ballistic behavior does not change 
but only does its subsequent diffusion speed [Fig. \ref{fig:diff_speed}(b)].
On the other hand, with a modification in the coin configuration,
the state $\ket{\varphi_-}$ shows dramatically different behaviors from  $\ket{\varphi_+}$.
First, while $\ket{\varphi_+}$ shows no appreciable difference between scalar and vector-type interactions,
$\ket{\varphi_-}$ gives very dissimilar ballistic behaviors depending on interaction types. 
That is, 
whereas much charge of the vector-type still resides at the origin,
most charge of the scalar-type travels away from the origin.
Second, related to this difference in spatial distribution,
there is a considerable distinction in the diffusion speed between the two interaction types.
Whereas the scalar-type walker exhibits an enhanced ballistic behavior, \ie, large value of $\sigma(t)/t$,
the vector type exhibits a suppressed ballistic behavior as its value of $\sigma(t)/t$ is even smaller than that of linear QW.
Notice that even though the fronts of charge---the leading part of traveling wave---almost coincide for scalar-type NQW and linear QW cases,
the solitonic feature enables the scalar type to have a larger ballistic speed.

For other evidence of wide range $\sigma(t)$,
we have picked 12,482 samples of initial states of the form
$\ket{x=0} \otimes [\cos (\theta/2) \ket{\uparrow} + \sin (\theta/2) e^{i\phi} \ket{\downarrow}]$ for
$\theta \in [0,\pi]$, $\phi \in [0,2\pi]$ with spacing 0.04 (rad) each and 
evaluated their ballistic diffusion speeds $\sigma(t)/t$ after 200 steps.
We plot in Fig. \ref{fig:diff_speed}(e) 
the values of $\sigma(t)/t$ for linear QW, scalar-type NQW, and vector-type NQW, respectively.
Here we have realigned the data sets in ascending order of $\sigma(t)/t$ of linear QW.
We see a drastic variation of the ballistic speed of the two interaction types of NQW.
Vector-type NQW has a larger speed than the linear QW in some cases 
but has an upper bound of $\sim$0.43 in the whole cases. In most cases, it thus has a smaller speed than linear QW.
In contrast, the scalar-type NQW shows a wide range of the speed values
so that just by changing an initial coin state 
it can have either a larger or smaller speed than each value of linear QW.

\section{Conclusion}
We showed that NQW is capable of simulating dynamics of nonlinear quantum models. We considered two kinds of nonlinear Dirac equations and observed solitonic behavior in both situations. Since NQW can be implemented in laboratories with present optical setups, our model provides a new tool to study solitons in physical systems. Moreover, formation of solitons leads to a completely new form of localization that has not been observed in quantum walk models before. Finally, we showed that nonlinearity offers a new method to control the spread of the quantum walker, making both enhancement and suppression of ballistic spreading possible. The control of the spread is achieved by a proper choice of the initial state of the coin degree of freedom.

Apart from the various applications mentioned above, nonlinear phenomena play a crucial role in the emergence of complexity \cite{Complex}. It is therefore important to investigate relatively simple models exhibiting such nonlinear behavior to achieve a deeper understanding of nonlinear physics. QWs are already known to be able to simulate various physical systems and it is natural to expect that the NQW model presented in this work can find applications in studies related to quantum complexity.

\section*{acknowledgments}
C.-W.L. enjoyed discussion with Fabian Maucher about nonlinear Schr\"odinger equation and its numerical solution.
C.-W.L. is supported by the IT R\&D program of MOTIE/KEIT [Grant No. 10043464(2012)].
P.K. acknowledges support from the National Research Foundation and Ministry of Education in Singapore. HN is supported by an NPRP Grant No. 7-210-1-032 from Qatar National Research Fund.


\setcounter{figure}{0}
\renewcommand{\thefigure}{A\arabic{figure}}


\begin{figure}[b]
	\topinset
	{\bfseries(a)}
	{\includegraphics[width=0.48\columnwidth]{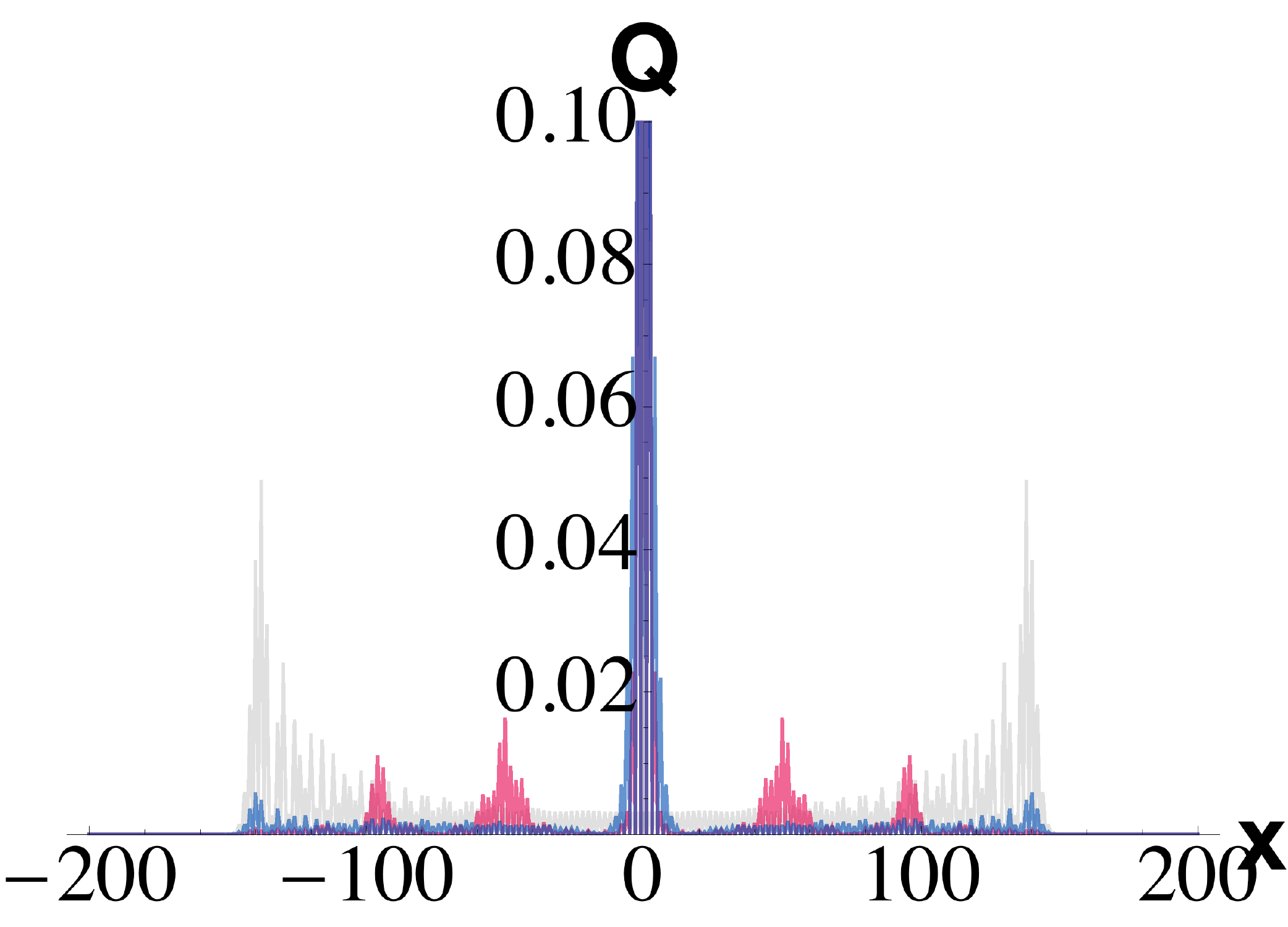}}
	{0mm}{35mm}
	\hskip2mm
	\topinset
	{\bfseries(b)}
	{\includegraphics[width=0.48\columnwidth]{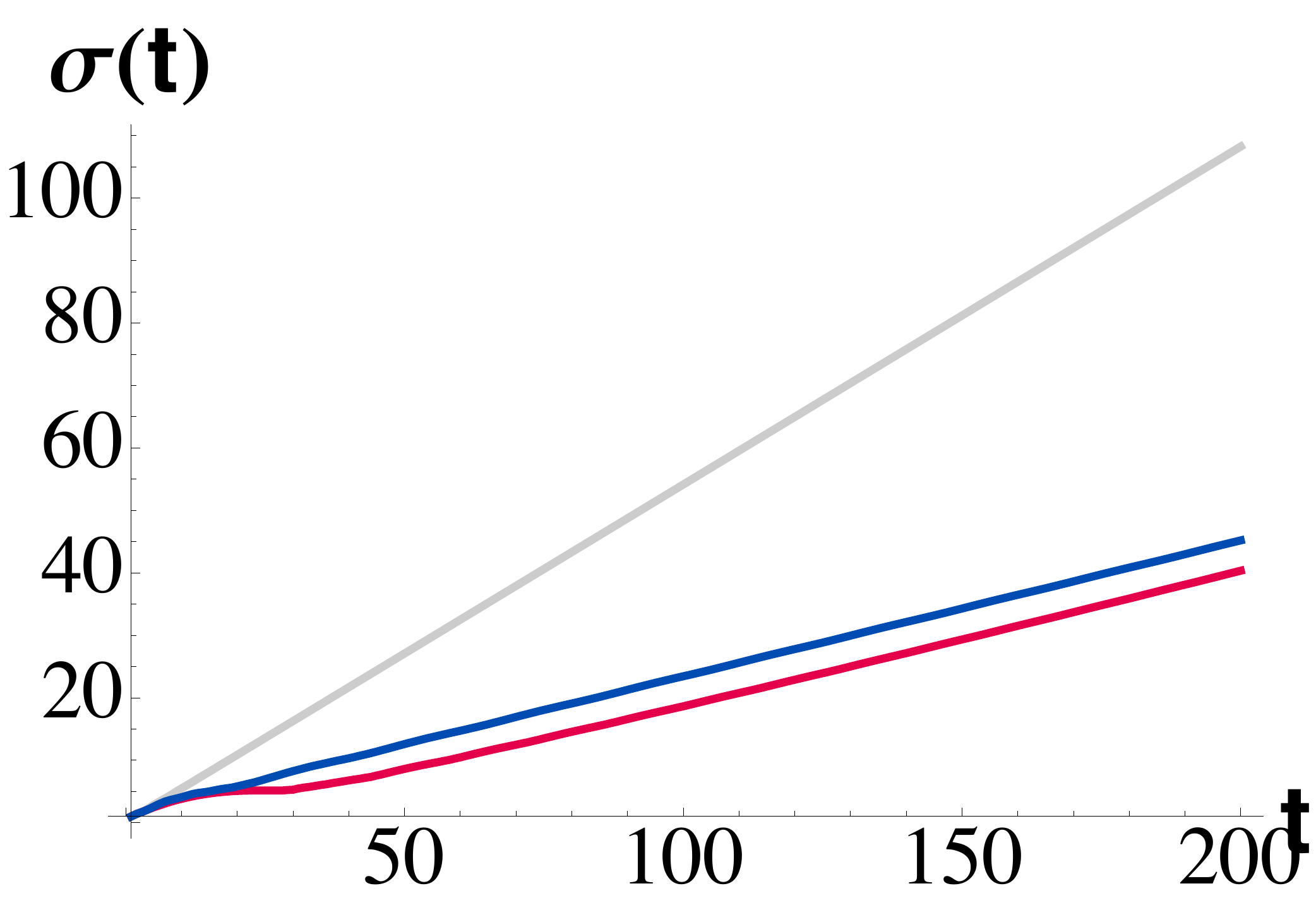}}
	{0mm}{35mm}
	\vskip4mm
	\topinset
	{\bfseries(c)}
	{\includegraphics[width=0.48\columnwidth]{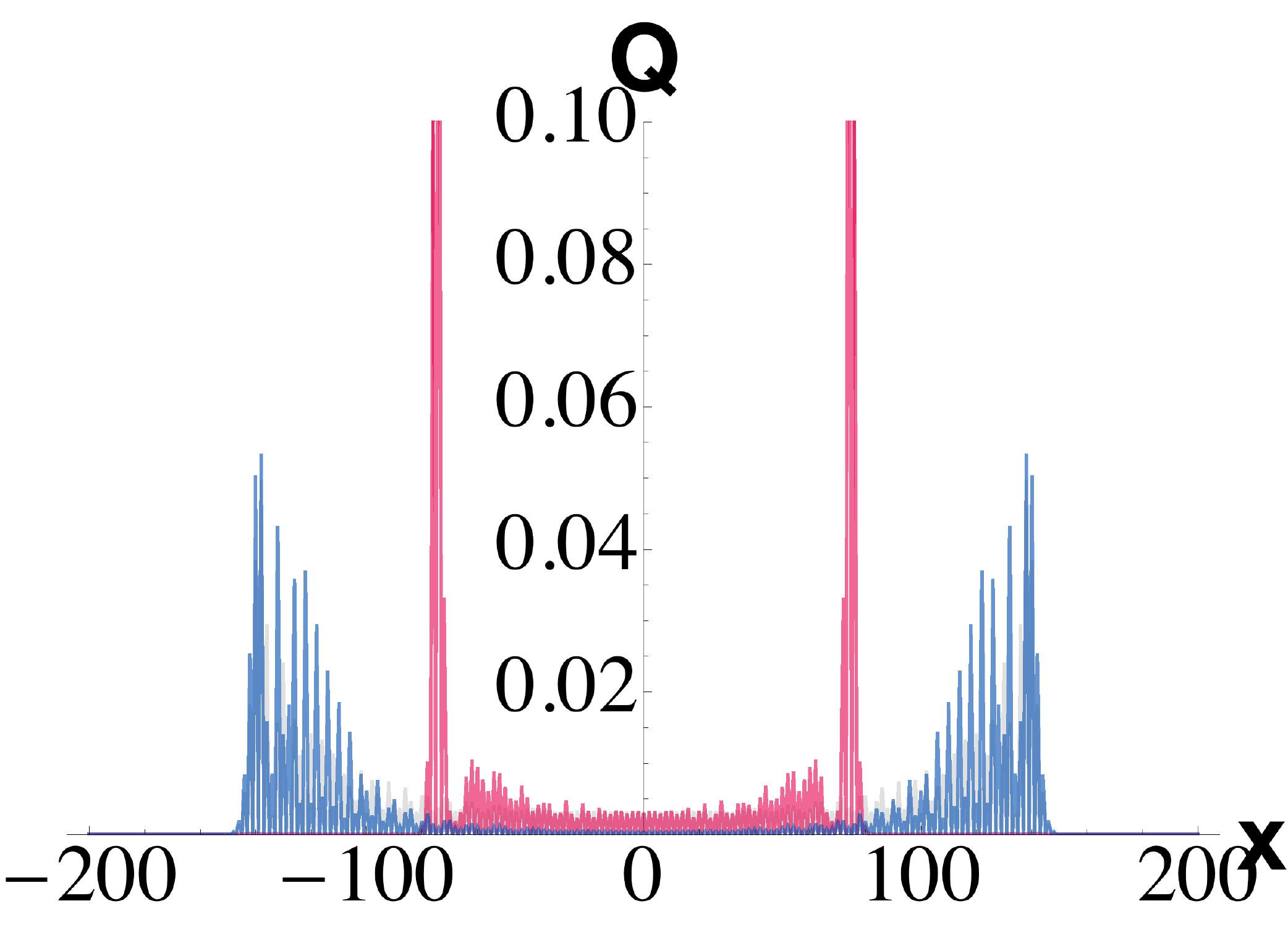}}
	{0mm}{35mm}
	\hskip2mm
	\topinset
	{\bfseries(d)}
	{\includegraphics[width=0.48\columnwidth]{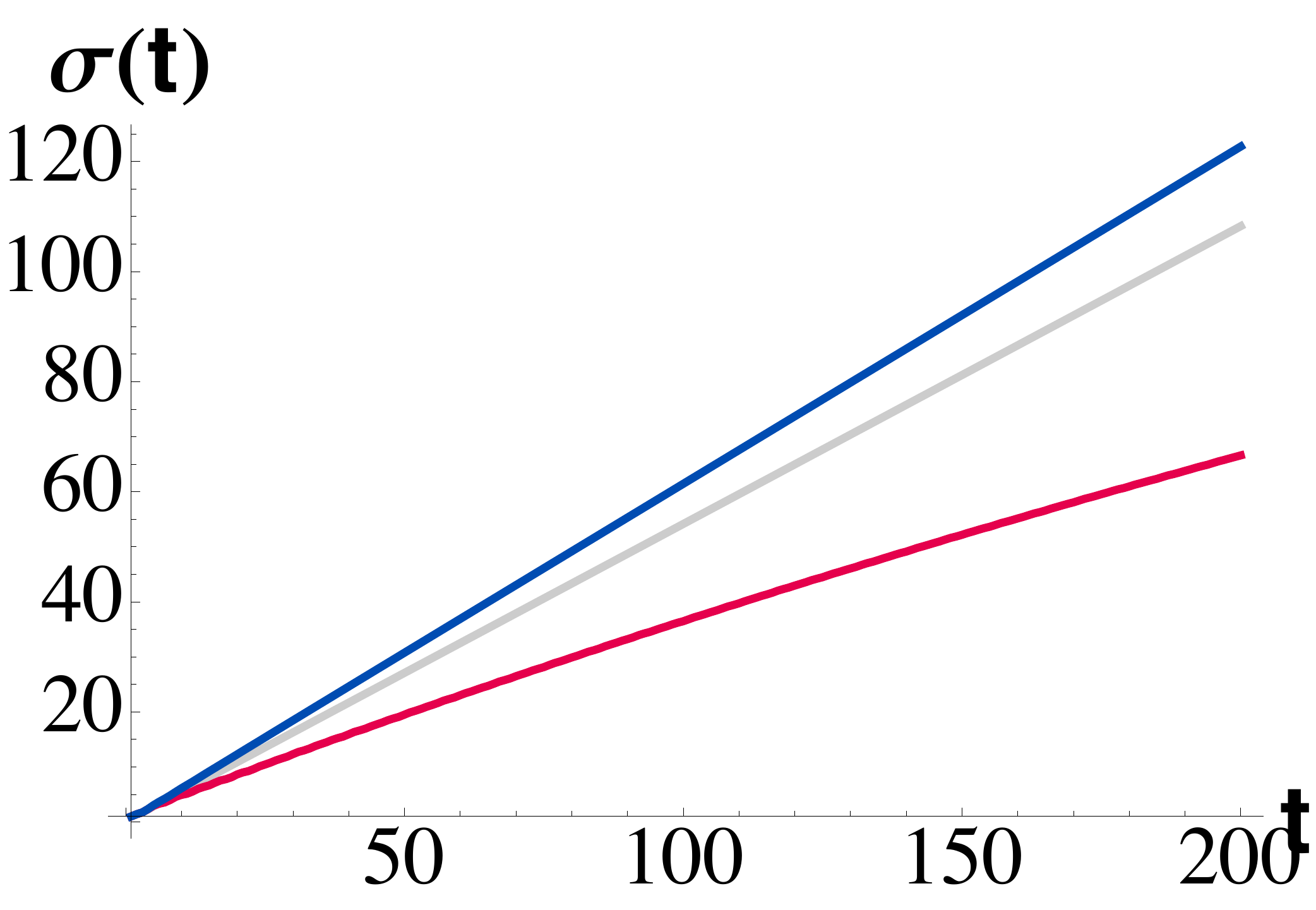}}
	{0mm}{35mm}
	\vskip4mm
	\topinset
	{\bfseries(e)}
	{\includegraphics[width=0.48\columnwidth]{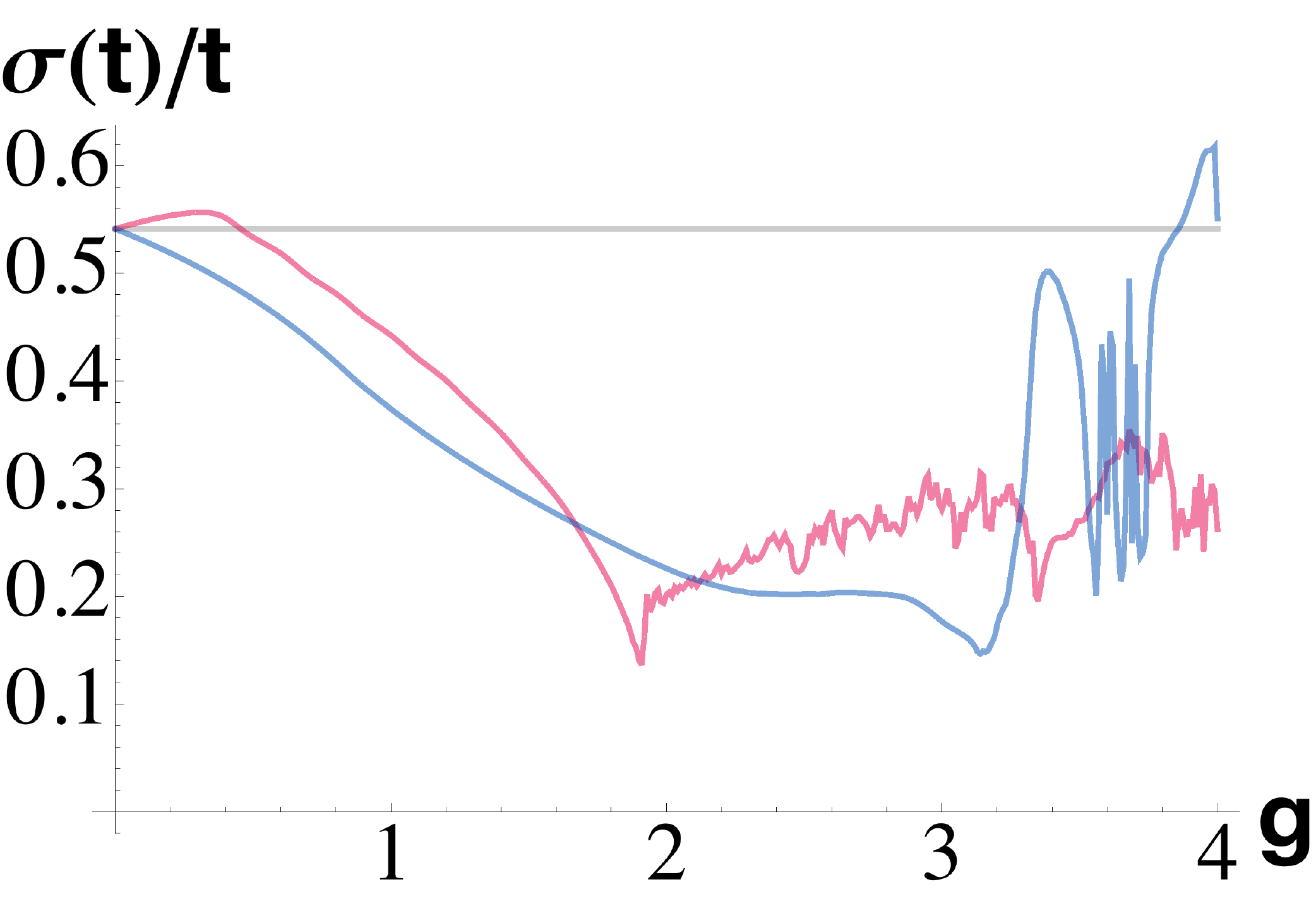}}
	{1mm}{35mm}
	\hskip2mm
	\topinset
	{\bfseries(f)}
	{\includegraphics[width=0.48\columnwidth]{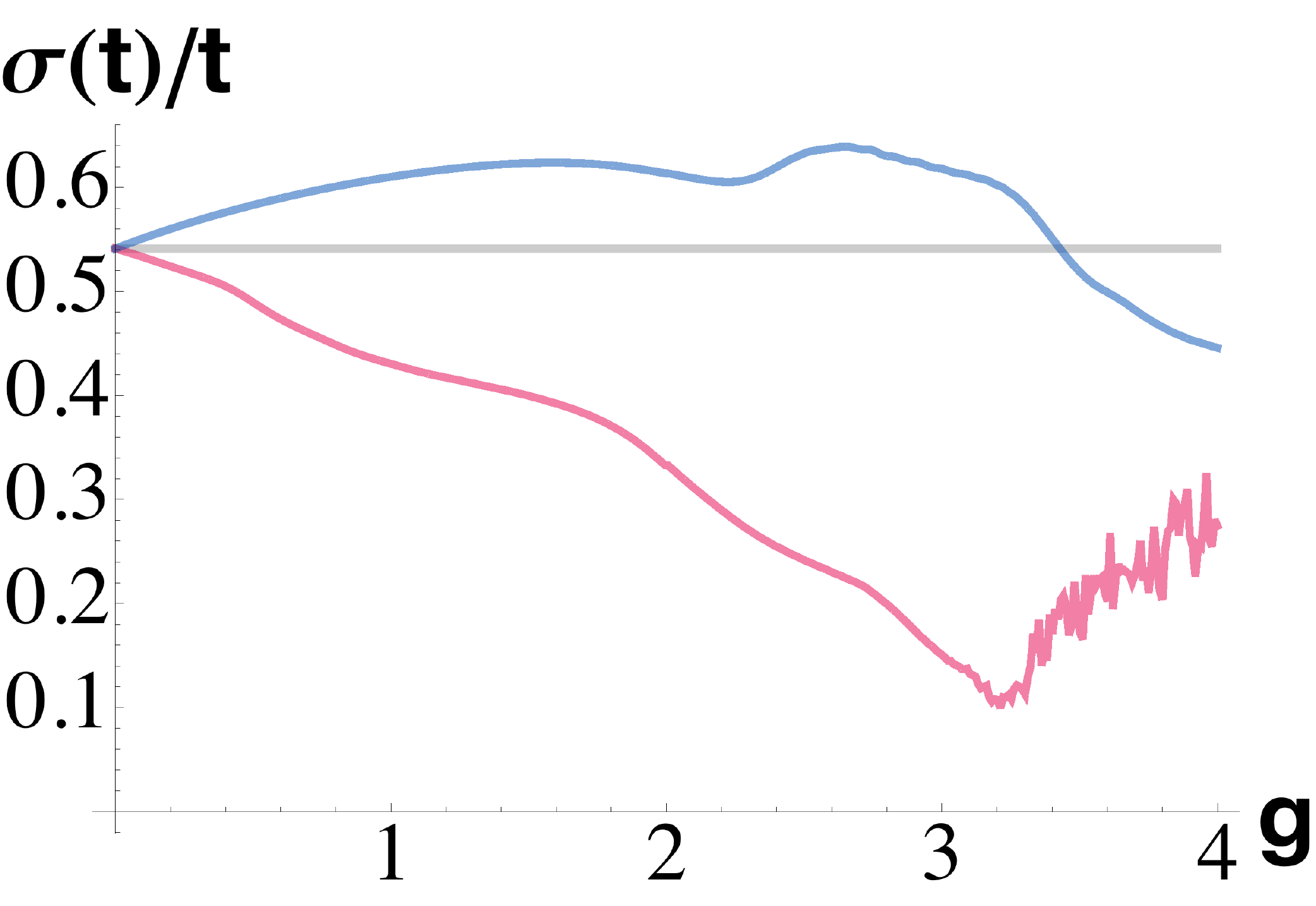}}
	{1mm}{35mm}
	\vskip4mm
	\caption{(Color online)
		Snapshot of the charge distribution at $t=200$ with the nonlinear-coupling strength $g=2$ (a) [(c)] and the corresponding growth of the position width $\sigma(t)$ (b) [(d)] for the case of $\ket{\varphi_+} = \ket{\uparrow} + i \ket{\downarrow}$ [$\ket{\varphi_-} = \ket{\uparrow} - i \ket{\downarrow}$]. 
		Ballistic diffusion speed $\sigma(t)/t$ as a function of coupling strength $g$ for $\ket{\varphi_+}$ (e) and for $\ket{\varphi_-}$ (f), respectively.
		In each plot the gray curve denotes the case of linear QW, the blue curve the case of scalar type, and the red curve the case of vector-type NQW.
	}
	\label{fig:diff_speed_g}
\end{figure}

\section*{Appendix: Influences of coupling constant and coin angle on ballistic diffusion speed}

\begin{figure}[h]
	\topinset
	{\bfseries(a)}
	{\includegraphics[width=0.48\columnwidth]{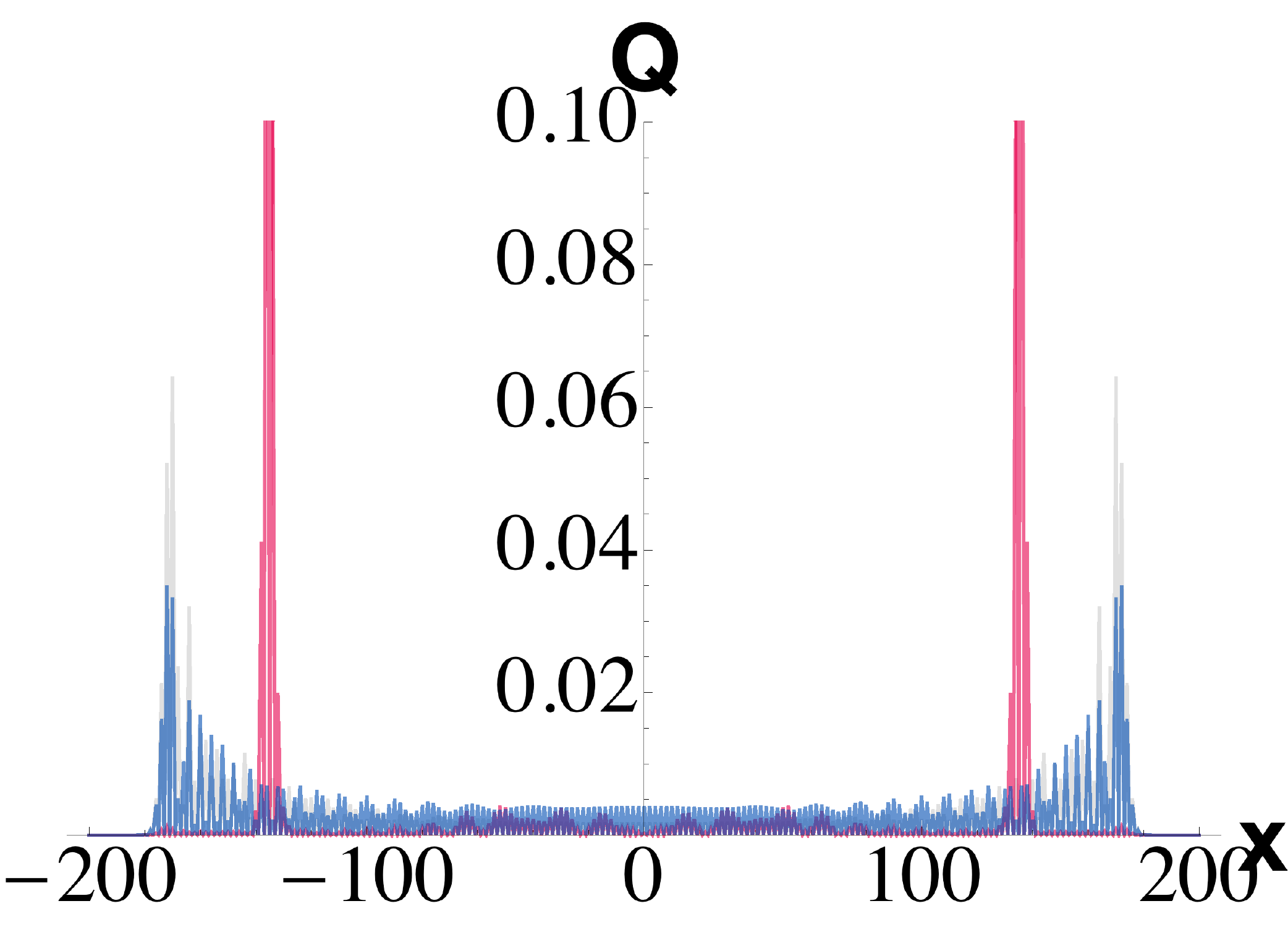}}
	{0mm}{35mm}
	\hskip2mm
	\topinset
	{\bfseries(b)}
	{\includegraphics[width=0.48\columnwidth]{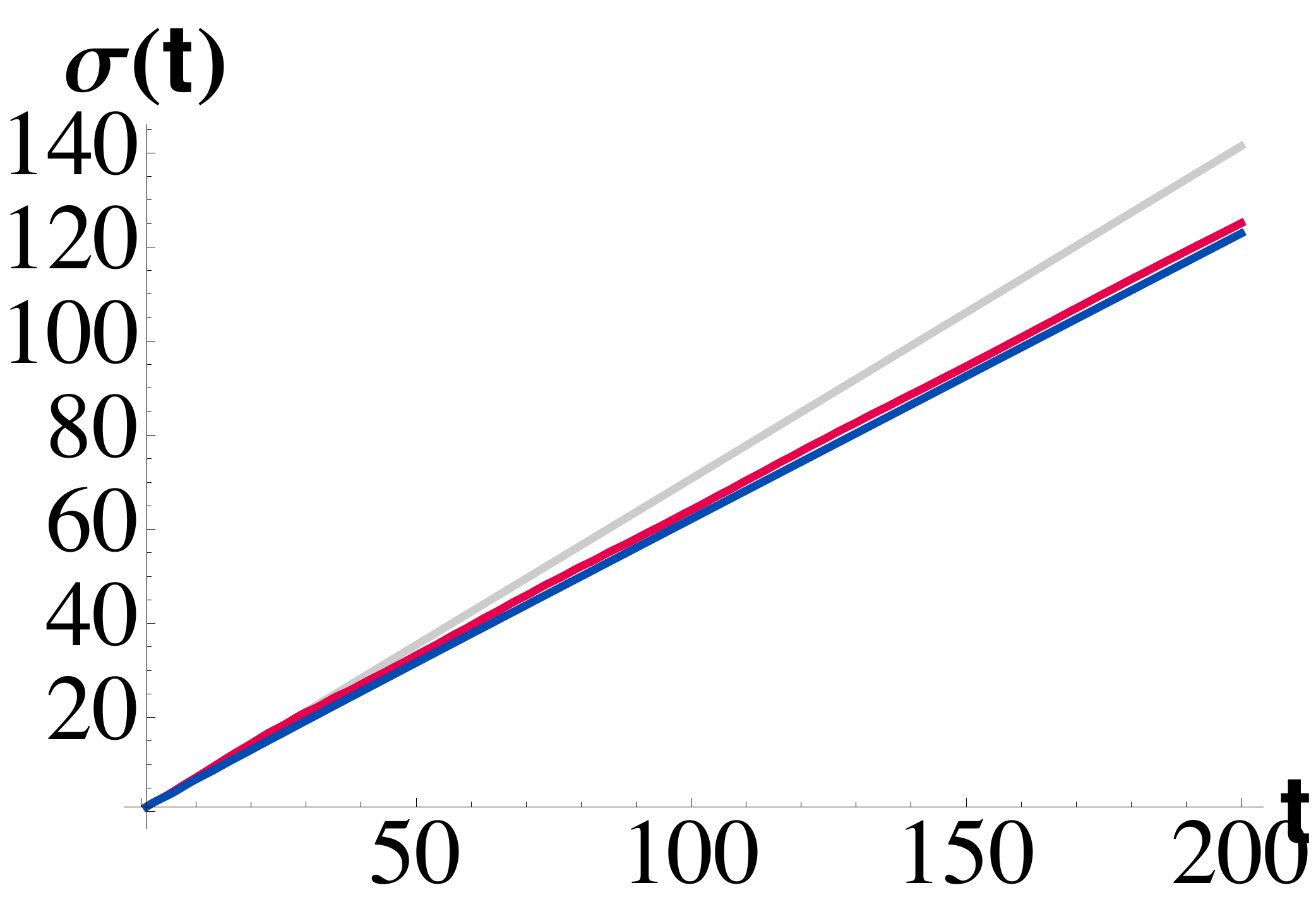}}
	{0mm}{35mm}
	\vskip4mm
	\topinset
	{\bfseries(c)}
	{\includegraphics[width=0.48\columnwidth]{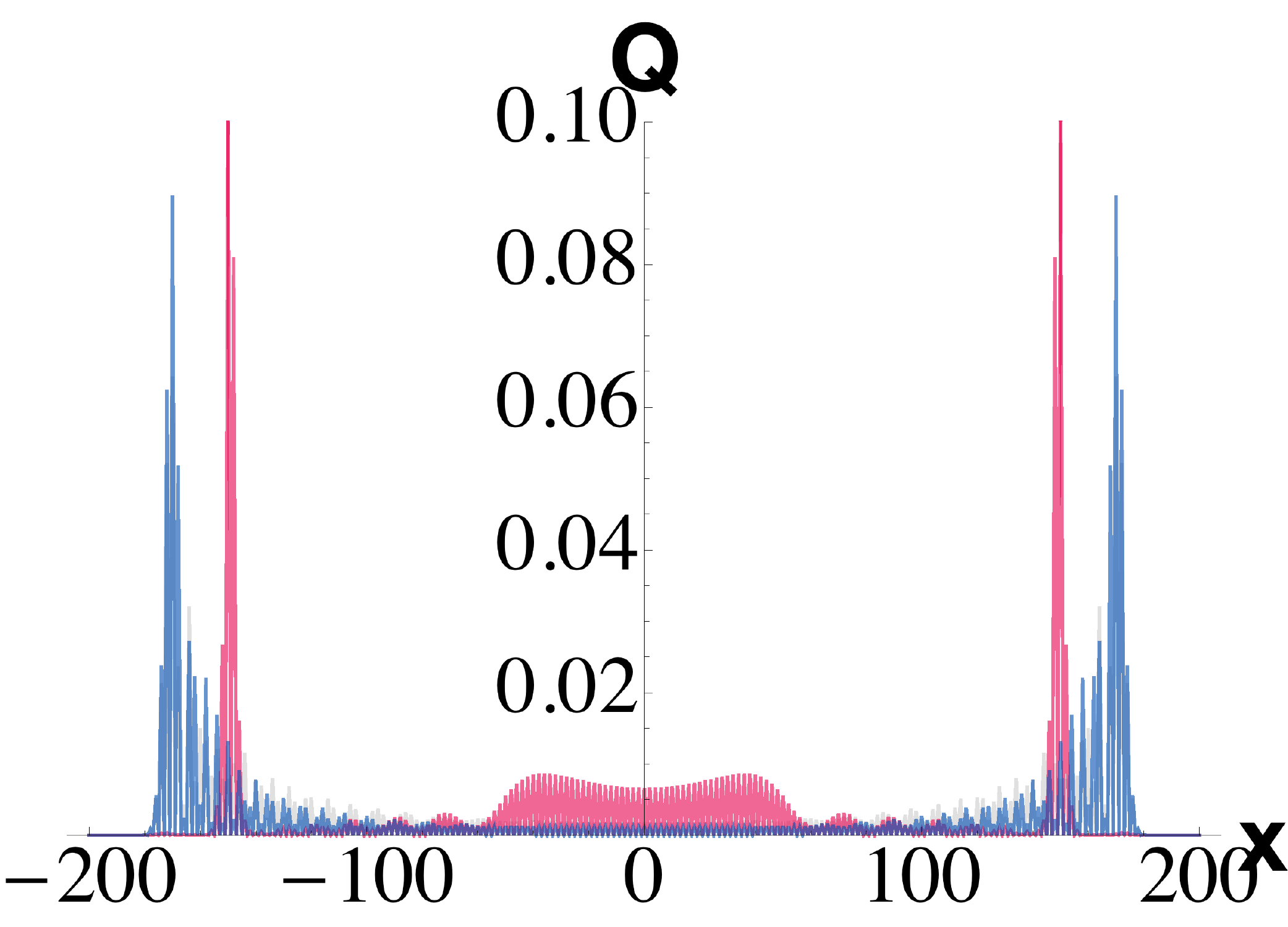}}
	{0mm}{35mm}
	\hskip2mm
	\topinset
	{\bfseries(d)}
	{\includegraphics[width=0.48\columnwidth]{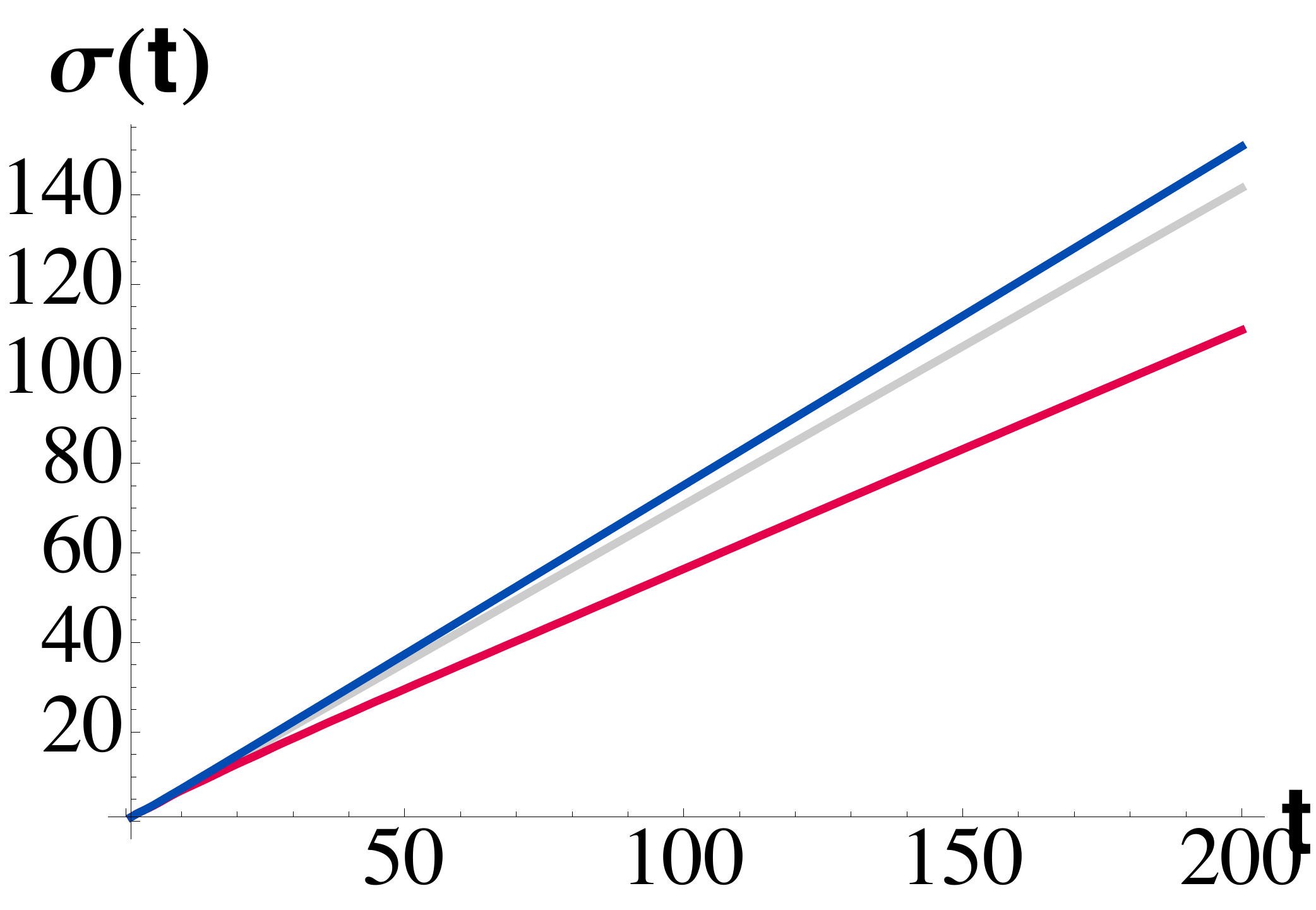}}
	{0mm}{35mm}
	\vskip4mm
	\topinset
	{\bfseries(e)}
	{\includegraphics[width=0.48\columnwidth]{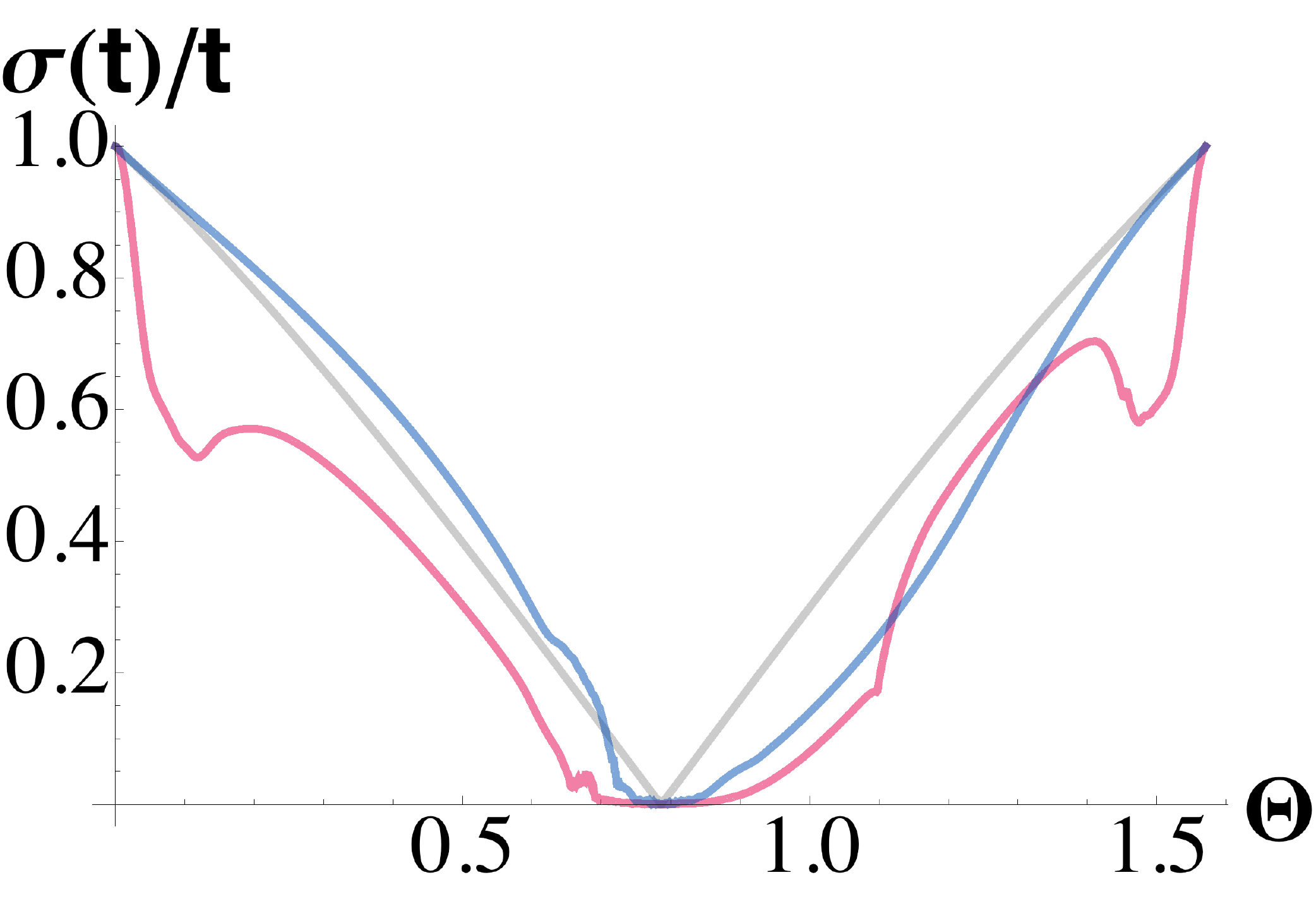}}
	{0mm}{35mm}
	\hskip2mm
	\topinset
	{\bfseries(f)}
	{\includegraphics[width=0.48\columnwidth]{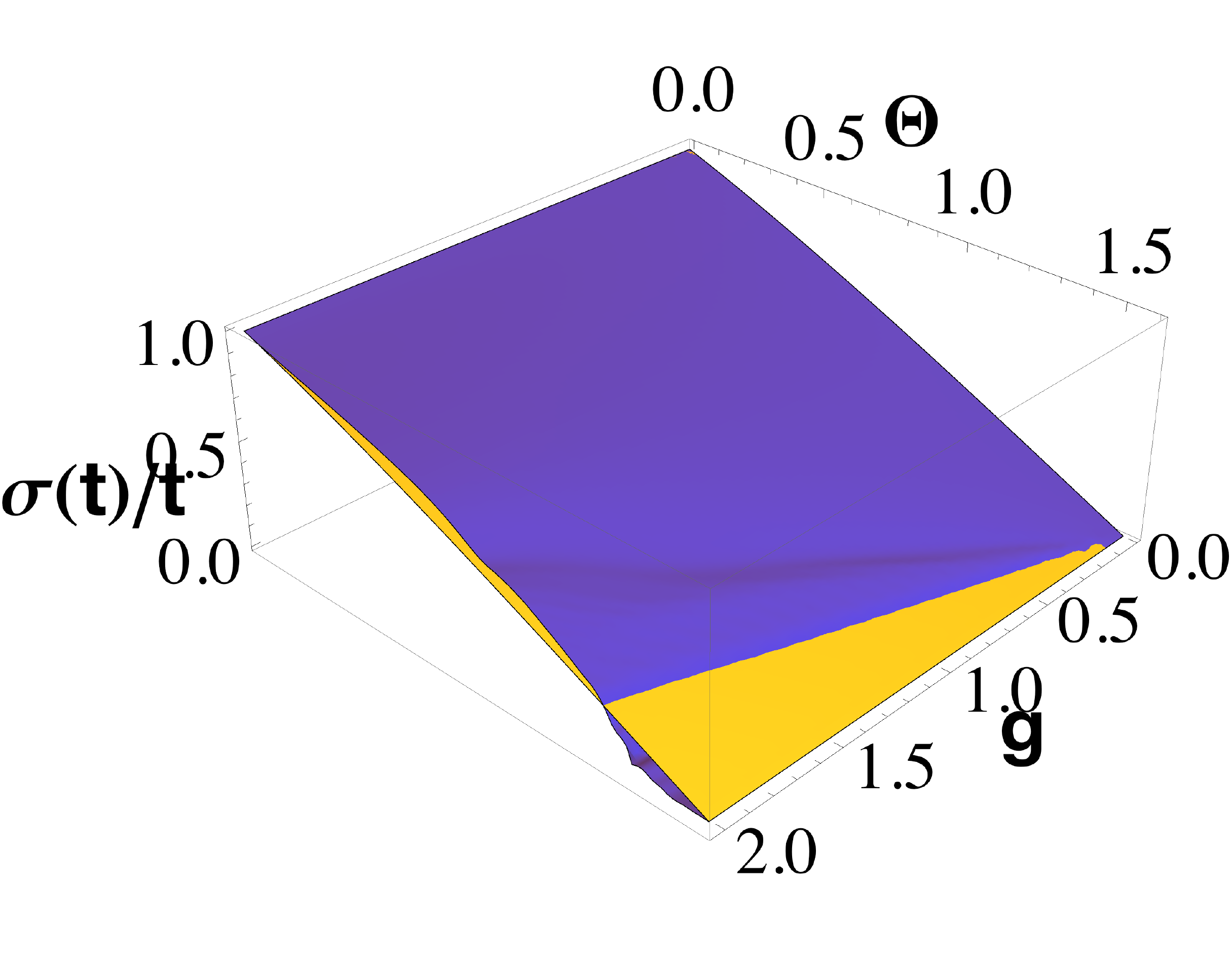}}
	{0mm}{35mm}
	\vskip4mm
	\caption{(Color online)
		Snapshot of the charge distribution at $t=200$ in the case of $\Theta=\pi/6$ (a) [(c)] 
		and the corresponding growth of the position width $\sigma(t)$ (b) [(d)] for the case of 
		$\ket{\varphi_+}$ [$\ket{\varphi_-}$]. 
		(e) Ballistic diffusion speeds of $\ket{\varphi_-}$ at $t=200$ as functions of $\Theta$.
		(f) Ballistic diffusion speeds of linear QW (yellow) and scalar-type case (violet) of $\ket{\varphi_-}$ as functions of coupling constant $g$ and coin angle $\theta$.
		Except for (f), the color legends of each curve are the same as in Fig. \ref{fig:diff_speed_g}.
	}
	\label{fig:diff_speed_theta}
\end{figure}

Although we have fixed the coupling constant $g$ and the coin angle $\Theta$ of constant step in the main text,
we can modify those parameters as well and see how these affect the ballistic diffusion speed $\sigma(t)/t$. 
We here examine this issue by some calculations, 
which are shown in Fig. \ref{fig:diff_speed_g} and \ref{fig:diff_speed_theta}.

First, 
we change the value of coupling constant from $g=1$ to $g=2$ and 
plot the charge distribution at $t=200$ and the position width $\sigma(t)$ for 
an initial state $\ket{\varphi_+} = \ket{\uparrow} + i \ket{\downarrow}$ in Figs. \ref{fig:diff_speed_g} (a) and (b)  
[$\ket{\varphi_-} = \ket{\uparrow} - i \ket{\downarrow}$ in Figs. \ref{fig:diff_speed_g} (c) and (d)], in comparison to Figs. 3 
(a)-(d).
Note that we used the same coin angle $\Theta= \pi/4$ of Eq. \eqref{eq:coin} 
in order to see how $g$ affects the behavior of $\sigma(t)$.
To see the effect of varying $g$ in more detail,
the values of $\sigma(t)/t$ at $t=200$ are computed for 
the state $\ket{\varphi_+}$ in Fig. \ref{fig:diff_speed_g} (e) and the state $\ket{\varphi_-}$ in Fig. \ref{fig:diff_speed_g} (f), respectively. 
For the case of $\ket{\varphi_+}$,
the diffusion speed $\sigma(t)/t$ tends to decrease with the nonlinear coupling strength $g$ for both type interactions, becoming smaller than that of linear QW
although the tendency ceases to be monotonic beyond certain values of $g$.
On the other hand, for the case of $\ket{\varphi_-}$, the vector-type case shows similar behavior to that of $\ket{\varphi_+}$
while the scalar type exhibits a larger value of $\sigma(t)/t$ than that of linear QW case.

Next,
we also check the effect of varying the coin angle $\Theta$ keeping $g=1$,
which accordingly determines the mass $m=\sin \Theta\cos \Theta$ of Dirac particle when the Hamiltonian of NQW is mapped to that of NDE.
We choose $\Theta= \pi/6$ again for $\ket{\varphi_\pm}$ and plot the result in Fig. \ref{fig:diff_speed_theta} (a)-(d).
The detailed dependence of $\sigma(t)/t$ on $\Theta$ for $\ket{\varphi_-}$ is shown in Fig. \ref{fig:diff_speed_theta} (e).
As can be seen from the plots, the ballistic diffusion speed $\sigma(t)/t$ of vector type is always smaller than that of linear QW
in the whole range of $\Theta$. Only the scalar type shows a larger value of $\sigma(t)/t$ in a narrow range of $\Theta$.
Note that the plot for $\ket{\varphi_+}$ is the same as Fig. \ref{fig:diff_speed_theta} (e) 
except that the curve is reflected about $\Theta=\pi/2$,
which is evident from the relation between the symmetries of $\ket{\varphi_\pm}$ and the coin operator in Eq. \eqref{eq:coin}. 
Furthermore, the ballistic diffusion speed of $\ket{\varphi_-}$ is plotted with respect to $g$ and $\Theta$ together in Fig. \ref{fig:diff_speed_theta} (f).
This shows that the value of $\Theta$ is more critical to $\sigma(t)/t$ than that of $g$ regarding the diffusion speed.

\end{document}